\documentclass[useAMS,usenatbib]{mn2e}
\usepackage{txfonts}
\usepackage{graphicx}
\usepackage{natbib}

\def\del#1{{}}

\newcounter{saveeqn}
\newcommand{\alpheqn}{
  \setcounter{saveeqn}{\value{equation}}%
  \addtocounter{saveeqn}{1}
  \setcounter{equation}{0}%
  \renewcommand{\theequation}{%
                \mbox{\arabic{section}.\arabic{saveeqn}\alph{equation}}}}
\newcommand{\reseteqn}{\setcounter{equation}{\value{saveeqn}}%
   \renewcommand{\theequation}{\arabic{section}.\arabic{equation}}}

\eqsecnum

\newcommand{\eps}{\varepsilon}
\newcommand{\crp}{\mathrm{CRp}}
\newcommand{\cre}{\mathrm{CRe}}
\newcommand{\e}{\mathrm{e}}
\newcommand{\p}{\mathrm{p}}
\newcommand{\NT}{\mathrm{NT}}
\newcommand{\hadr}{\mathrm{hadr}}
\newcommand{\dd}{\mathrm{d}}

\title[Minimum energy criteria] {Estimating galaxy cluster magnetic fields by
  the classical and hadronic minimum energy criterion}

\author[C. Pfrommer and T. A. En{\ss}lin] {C. Pfrommer\thanks{e-mail:
    pfrommer@mpa-garching.mpg.de (CP); ensslin@mpa-garching.mpg.de (TAE)} and
  T. A. En{\ss}lin\footnotemark[1]\\
  Max-Planck-Institut f\"ur Astrophysik, Karl-Schwarzschild-Stra{\ss}e 1,
  Postfach 1317, 85741 Garching, Germany}

\begin{document}
\pagerange{\pageref{firstpage}--\pageref{lastpage}} \pubyear{2003}
\maketitle
\label{firstpage}

% --- abstract --- %
\begin{abstract}
  We wish to estimate magnetic field strengths of radio emitting galaxy
  clusters by minimising the non-thermal energy density contained in cosmic ray
  electrons (CRe), protons (CRp), and magnetic fields. The {\em classical}
  minimum energy estimate can be constructed independently of the origin of the
  radio synchrotron emitting CRe yielding thus an absolute minimum of the
  non-thermal energy density.  Provided the observed synchrotron emission is
  generated by a CRe population originating from hadronic interactions of CRp
  with the ambient thermal gas of the intra-cluster medium, the parameter space
  of the {\em classical} scenario can be tightened by means of the {\em
    hadronic} minimum energy criterion.  For both approaches, we derive the
  theoretically expected tolerance regions for the inferred minimum energy
  densities.  Application to the radio halo of the Coma cluster and the radio
  mini-halo of the Perseus cluster yields equipartition between cosmic rays and
  magnetic fields within the expected tolerance regions. In the hadronic
  scenario, the inferred central magnetic field strength ranges from
  $2.4~\umu\rmn{G}$ (Coma) to $8.8~\umu\rmn{G}$ (Perseus), while the optimal
  CRp energy density is constrained to $2\% \pm 1\%$ of the thermal energy
  density (Perseus). We discuss the possibility of a hadronic origin of the
  Coma radio halo while current observations favour such a scenario for the
  Perseus radio mini-halo.  Combining future expected detections of radio
  synchrotron, hard X-ray inverse Compton, and hadronically induced
  $\gamma$-ray emission should allow an estimate of volume averaged cluster
  magnetic fields and provide information about their dynamical state.
\end{abstract}

% --- keywords --- %

% --- section: introduction --- %
\section{Introduction}
\label{sec:intro}

Clusters of galaxies harbour magnetised plasma. In particular, the detection of
diffuse synchrotron radiation from radio halos or relics provides evidence for
the existence of magnetic fields within the intra-cluster medium (ICM)
\citep[for a review, see][]{2002ARA&A..40..319C}. Since the detection rate of
radio halos in galaxy clusters seems to be of the order of 30\% for X-ray
luminous clusters \citep{1999NewA....4..141G}, the presence of magnetic fields
appears to be common.  Based on these observations, \citet{2002A&A...396...83E}
developed a redshift dependent radio halo luminosity function and predicted
large numbers of radio halos to be detected with future radio telescopes.

A different piece of evidence comes from Faraday rotation which arises owing to
the birefringent property of magnetised plasma causing the plane of
polarisation to rotate for a nonzero magnetic field component along the
propagation direction of the photons \citep{2001ApJ...547L.111C}.  However, the
accessible finite windows given by the extent of the sources emitting polarised
radiation are a limitation of this method.  The derived magnetic field
strengths depend on the unknown magnetic field autocorrelation length which has
to be deprojected from the observed two dimensional Faraday rotation measure
maps using certain assumptions \citep[see, however,][]{2003A&A...401..835E,
  2003A&A...412..373V}.  A different approach is given by the energy
equipartition argument if a particular cluster exhibits diffuse radio
synchrotron emission.  The method assumes equal energy densities of cosmic ray
electrons and magnetic fields in order to estimate volume averaged magnetic
field strengths.

The minimum energy criterion is a complementary method.  It is based on the
idea of minimising the non-thermal energy density contained in cosmic ray
electrons (CRe), protons (CRp), and magnetic fields by varying the magnetic
field strength. As one boundary condition, the implied synchrotron emissivity
is required to match the observed value. Additionally, a second boundary
condition is required mathematically which couples CRp and CRe.  For the
classical case, a constant scaling factor between CRp and CRe energy densities
is assumed. However, if the physical connection between CRp and CRe is known or
assumed, a physically better motivated criterion can be formulated.  As such a
case, we introduce the minimum energy criterion within the scenario of
hadronically generated CRe.

Classically, the equipartition/minimum energy formulae use a fixed interval in
radio frequency in order to estimate the total energy density in cosmic ray
electrons (CRe), a purely observationally motivated procedure
\citep{1956ApJ...124..416B, 1970ranp.book.....P}. However, this approach has a
drawback when comparing different field strengths between galaxy clusters
because a given frequency interval corresponds to different CRe energy
intervals depending on the magnetic field strengths
\citep{2001SSRv...99..243B}.  For this reason, variants of the minimum energy
criterion have been studied in order to place the magnetic field estimates on
more physical grounds, based then on assumptions such as the fixed interval in
CRe energy \citep{1993A&A...270...91P, 1996ARA&A..34..155B,
  1997A&A...325..898B}. The modified classical minimum energy criterion does
not specify a particular energy reservoir of the CRe.  However, this apparent
model-independence is bought dearly at the cost of the inferred magnetic field
strength depending on unknown parameters like the lower energy cutoff of the
CRe population or the unknown contribution of CRp to the non-thermal energy
density. In the following, we use the term {\em classical minimum energy
  criterion} in its modified version, including e.g.~a fixed interval in CRe
energy.

Natural candidates for acceleration mechanisms providing a highly-relativistic
particle population are strong structure formation and merger shocks
\citep[e.g.,][]{1980A&AS...39..215H, 1999ApJ...520..529S} or reacceleration
processes of 'mildly' relativistic CRe ($\gamma_\e\simeq 100-300$) being
injected over cosmological timescales into the ICM. Owing to their long
lifetimes of a few times $10^9$ years, these mildly relativistic CRe can
accumulate within the ICM \citep[see][ and references
therein]{2002mpgc.book....1S}, until they experience continuous in-situ
acceleration via resonant pitch angle scattering by turbulent Alfv\'en waves as
originally proposed by \citet{1977ApJ...212....1J} and reconsidered by
\citet{1987A&A...182...21S}, \citet{2001MNRAS.320..365B},
\citet{2002ApJ...577..658O}, \citet{2002A&A...386..456G}, and
\citet{2003ApJ...594..732K}. However, this reacceleration scenario also faces
challenges as recent results imply: \citet{2003astro.ph.12482B} show, that if
the CRp-to-thermal energy density ratio were more than a few percent, Alfv\'en
waves would be damped efficiently such that the reacceleration mechanism of the
electrons is inefficient.  Because nearly all conceivable electron acceleration
mechanisms produce a population of CRp which accumulates within the clusters
volume, this represents an efficient damping source of Alfv\'en
waves.\footnote{Indeed, there are first hints for the existence of a 10~MeV -
  100~MeV CRp population deriving from the detection of excited gamma-ray lines
  from the clusters Coma and Virgo \citep{2004A&A.413..817I}. If verified, that
  would make a high energy (GeV) CRp population very plausible.}
\citet{2004ApJ...604..108K} presented an interesting line of argumentation to
investigate the nature of radio halos by comparing the observed and
statistically predicted population. This approach might allow to measure the
life time of radio halos and thus help to conclude their physical origin with a
future flux-limited, controlled, and homogeneous radio halo sample.

In this work, we examine a minimum energy criterion within another specific
model for the observed extended radio halos of $\sim$~Mpc size: Hadronic
interactions of CRp with the ambient thermal gas produce secondary electrons,
neutrinos, and $\gamma$-rays by inelastic collisions taking place throughout
the cluster volume.  These secondary CRe would generate radio halos through
synchrotron emission \citep{1980ApJ...239L..93D, 1982AJ.....87.1266V,
  1999APh....12..169B, 2000A&A...362..151D, 2001ApJ...559...59M,
  2004A&A...413...17P}. This scenario is motivated by the following argument:
The radiative lifetime of a CRe population in the ICM, generated by direct
shock acceleration, is of the order of $10^8$ years for $\gamma_\e \sim 10^4$.
This is relatively short compared to the required diffusion timescale needed to
account for such extended radio phenomena \citep{2002BrunettiTaiwan}.  On the
other hand, the CRp are characterised by lifetimes of the order of the Hubble
time, which is long enough to diffuse away from the production site and to be
distributed throughout the cluster volume to which they are confined by
magnetic fields \citep{1996SSRv...75..279V, 1997ApJ...477..560E,
  1997ApJ...487..529B}.  The magnetic field strength within this scenario is
obtained by analogy with the classical minimum energy criterion while combining
the CRp and CRe energy densities through their physically connecting process.
Apart from relying on the particular model, the inferred magnetic field
strengths do not depend strongly on unknown parameters in this
model.\footnote{Likewise the minimum energy criterion within the reacceleration
  scenario of mildly relativistic CRe ($\gamma_\e\simeq 100-300$) can be
  obtained by minimising the sum of magnetic, mildly relativistic CRe, and
  turbulent energy densities while allowing for constant synchrotron emission.}

The philosophy of these approaches is to provide a criterion for the
energetically least expensive radio synchrotron emission model possible for a
given physically motivated scenario.  To our knowledge, there is no first
principle enforcing this minimum state to be realised in Nature. However, our
minimum energy estimates are interesting in two respects: First, these
estimates allow scrutinising the hadronic model for extended radio synchrotron
emission in clusters of galaxies.  If it turns out that the required minimum
non-thermal energy densities are too large compared to the thermal energy
density, the hadronic scenario will become implausible to account for the
extended diffuse radio emission. For the classical minimum energy estimate,
such a comparison can yield constraints on the accessible parameter space
spanned by lower energy cutoff of the CRe population or the contribution of CRp
to the non-thermal energy density.  Secondly, should the hadronic scenario be
confirmed, the minimum energy estimates allow testing for the realisation of
the minimum energy state for a given independent measurement of the magnetic
field strength.

This article is organised as follows: After introducing synchrotron radiation
of CRe (Sect.~\ref{sec:CRe}), analytic formulae for hadronically induced
emission processes are presented (Sect.~\ref{sec:CRp}). The classical and
hadronic minimum energy criteria are then derived, the theoretically expected
tolerance regions are given, and limiting cases are discussed (Sect.~\ref{sec:MEC}).
In Sect.~\ref{sec:testing}, we examine whether future observations of inverse
Compton emission and hadronically induced $\gamma$-ray emission can serve as
tests for the verification of the minimum energy criterion. Magnetic and cosmic
ray energy densities and their tolerance regions are inferred from application of the
minimum energy arguments in the Coma and Perseus cluster for both scenarios
(Sect.~\ref{sec:applications}).  Our article concludes with formulae which
provide recipes for estimating the magnetic field strength in typical
observational situations (Sect.~\ref{sec:nutshell}).  Throughout this paper we
use the present Hubble constant $H_0 = 70~h_{70}\mbox{ km s}^{-1} \mbox{
  Mpc}^{-1}$, where $h_{70}$ indicates the scaling of $H_0$.

% --- section: basics --- %
\section{Theoretical background}
\label{sec:basics}
This section presents our definitions and the theoretical background for this
work. After introducing characteristics of the CRe population and the
synchrotron emission formulae, we focus on specifications of the CRp
population.  Finally, the section concludes with analytic formulae describing
the hadronically induced $\gamma$-ray and radio synchrotron emission processes.

\subsection{Cosmic ray electrons and synchrotron emission}
\label{sec:CRe}

The differential number density distribution of a CRe population above a MeV is
often represented by a power-law in energy $E_\e$,
\begin{equation}
\label{fe}
f_\e (\bmath{r}, E_\e) \,\dd E_\e\,\dd V = 
\tilde{n}_\cre(\bmath{r})\, \left(\frac{E_\e}{\mbox{GeV}} \right)^{-\alpha_\e}\,
\left(\frac{\dd E_\e}{\mbox{GeV}}\right)\, \dd V.
\end{equation}
where the tilde indicates that $\tilde{n}_\cre$ is not a real CRe number
density although it exhibits the appropriate dimensions.  The normalisation
$\tilde{n}_\cre(\bmath{r})$ might be determined by assuming that the kinetic
CRe energy density $\eps_{\cre} (\bmath{r})$ is expressed in terms of the
thermal energy density $\eps_\rmn{th} (\bmath{r})$,
\begin{eqnarray}
\label{Xcrescaling}
\eps_\cre(\bmath{r}) &=&  X_\cre(\bmath{r}) \, \eps_\rmn{th}(\bmath{r}) 
= A_{E_\e}\, \tilde{n}_\cre(\bmath{r}) , \\ 
\label{AEe}
A_{E_\e}(\alpha_\e) &=&   \frac{\mbox{GeV}}{2-\alpha_\e}
\left[\left(\frac{E_\e}{\mbox{GeV}}\right)^{2-\alpha_\e}\right]_{E_1}^{E_2}.
\end{eqnarray}
Here we introduced the abbreviation
\begin{equation}
[f(x)]_{x_1}^{x_2} = f(x_2) - f(x_1)
\end{equation}
in order to account for cutoffs of the CRe population.  If the CRe population
had time to lose energy by means of Coulomb interactions
\citep{1972Physica....58..379G}, the low energy part of the spectrum would be
modified. This modification, which impacts on the CRe distribution function
$f_\e (\bmath{r}, E_\e)$ and thus on $A_{E_\e}(\alpha_\e)$, can be
approximately treated by imposing a time dependent lower energy cutoff on the
CRe population as described in \citet{2004A&A...413...17P}.

While the functional dependence of the CRe scaling parameter
$X_\cre(\bmath{r})$ is a priori unknown, its radial behaviour will be adjusted
such that it obeys the minimum energy criterion.  The thermal energy density
of the ICM $\eps_\rmn{th}$ is given by
\begin{eqnarray}
\eps_\rmn{th} (\bmath{r}) &=& \frac{3}{2}\,d_\e\,n_\e(\bmath{r})\,k
T_\e(\bmath{r}),\label{eps thermal}\\
\mbox{where}\quad
d_\e &=& 1 + \frac{1 - \frac{3}{4} X_\rmn{He}}
{1 - \frac{1}{2} X_\rmn{He}}
\end{eqnarray}
counts the number of particles per electron in the ICM using the primordial
${}^4\rmn{He}$ mass fraction $X_\rmn{He} = 0.24$. $T_\e$ and $n_\e$ denote the
electron temperature and number density, respectively.

The synchrotron emissivity $j_\nu$ at frequency $\nu$ and per steradian of such
a CRe population (\ref{fe}), which is located in an isotropic distribution of
magnetic fields \citep[Eq.~(6.36) in][]{1979rpa..book.....R}, is obtained after
averaging over an isotropic distribution of electron pitch angles yielding
\begin{eqnarray}
\label{jnu}
j_\nu(\bmath{r})&=& A_{E_\rmn{syn}}(\alpha_\e)\, \tilde{n}_\cre(\bmath{r})\,
\left[\frac{\eps_B(\bmath{r})}{\eps_{B_\rmn{c}}}\right]^{(\alpha_\nu+1)/2}
\\
&\propto& \eps_\cre(\bmath{r})\, B(\bmath{r})^{\alpha_\nu+1} \nu^{-\alpha_\nu},
\\
\label{Bc}
B_\rmn{c} &=& \sqrt{8\upi\, \eps_{B_\rmn{c}}} 
   = \frac{2\upi\, m_\e^3\,c^5\, \nu}{3\,e \mbox{ GeV}^2}
   \simeq 31 \left(\frac{\nu}{\mbox{GHz}} \right)~\umu\rmn{G},
\\
A_{E_\rmn{syn}} &=& \frac{\sqrt{3\upi}}{32 \upi}
   \frac{B_\rmn{c}\, e^3}{m_\e c^2}
   \frac{\alpha_\e + \frac{7}{3}}{\alpha_\e + 1}
   \frac{\Gamma\left( \frac{3\alpha_\e-1}{12}\right)
         \Gamma\left( \frac{3\alpha_\e+7}{12}\right)
         \Gamma\left( \frac{\alpha_\e+5}{4}\right)}
        {\Gamma\left( \frac{\alpha_\e+7}{4}\right)},
\end{eqnarray}
where $\Gamma(a)$ denotes the Gamma-function \citep{1965hmfw.book.....A} and
$\alpha_\nu = (\alpha_\e-1)/2$.  Note that for later convenience, we introduce
a (frequency dependent) characteristic magnetic field strength $B_\rmn{c}$
which implies a characteristic magnetic energy density $\eps_{B_\rmn{c}}$.
Line-of-sight integration of the radio emissivity $j_\nu (\bmath{r})$ yields
the surface brightness of the radio emission $S_\nu(\bmath{r}_\bot)$.

\subsection{Cosmic ray protons}
\label{sec:CRp}

\subsubsection{CRp population}

In contrast to the previously introduced CRe population and owing to the higher
rest mass of protons, we assume the differential number density distribution of
a CRp population to be described by a power-law in momentum $p_\p$ which for
instance is motivated by shock acceleration studies:
\begin{equation}
\label{fp}
f_\p (\bmath{r}, p_\p) \,\dd p_\p\,\dd V = \tilde{n}_\crp(\bmath{r})\, 
\left(\frac{p_\p \,c}{\mbox{GeV}} \right)^{-\alpha_\p}\,
\left(\frac{c\,\dd p_\p}{\mbox{GeV}}\right)\, \dd V.
\end{equation}
The normalisation $\tilde{n}_\crp(\bmath{r})$ can be determined in such a way
that the kinetic CRp energy density $\eps_{\crp} (\bmath{r})$ is expressed in
terms of the thermal energy density $\eps_\rmn{th} (\bmath{r})$ of the ICM,
\begin{eqnarray}
\label{Xcrpscaling}
\eps_\crp(\bmath{r}) &=&  X_\crp(\bmath{r}) \, \eps_\rmn{th}(\bmath{r}) = 
A_{E_\p}\, \tilde{n}_\crp(\bmath{r}) , \\ 
\label{AEp}
A_{E_\p}(\alpha_\p) &=&   \frac{m_\p \,c^2}{2\,(\alpha_\p-1)}\,
\left(\frac{m_\p \,c^2}{\mbox{GeV}}\right)^{1-\alpha_\p}
\mathcal{B}\left(\frac{\alpha_\p-2}{2},\frac{3-\alpha_\p}{2}\right).
\end{eqnarray}
$\mathcal{B}(a,b)$ denotes the Beta-function \citep{1965hmfw.book.....A}.

Aging imprints a modulation on the low energy part of the CRp spectrum by
Coulomb losses in the plasma. This modification, which impacts on the CRp
distribution function $f_\p (\bmath{r}, p_\p)$ and thus on
$A_{E_\p}(\alpha_\p)$, can be treated approximately by imposing a lower
momentum cutoff as described in \citet{2004A&A...413...17P}.  On the other
side, highly energetic CRp with energies beyond $2\times 10^{7} \mbox{ GeV}$
are able to escape from the galaxy cluster assuming momentum dependent CRp
diffusion in a turbulent magnetic field with a Kolmogorov-type spectrum on
small scales \citep{1997ApJ...487..529B}. The finite lifetime and size of
particle accelerating shocks also give rise to high-energy breaks in the CRp
spectrum. These low and high momentum cutoffs are always present in the
CRp population. However, for CRp spectral indices between
$2\lesssim\alpha_\p\lesssim 3$ these spectral breaks have negligible influence
on the CRp energy density. If the breaks are neglected, the CRp energy density
would diverge for $\alpha_\p\lesssim 2$ at the high-energy and for
$3\lesssim\alpha_\p$ at the low-energy part of the spectrum. In these cases,
breaks have to be included by replacing $A_{E_\p}$ in Eq.~(\ref{Xcrpscaling})
with
\begin{eqnarray}
\label{cutoffs}
\tilde{A}_{E_\p}(\alpha_\p) &=&   \frac{m_\p \,c^2}{2\,(\alpha_\p-1)}\,
\left(\frac{m_\p \,c^2}{\mbox{GeV}}\right)^{1-\alpha_\p}
\left[ \mathcal{B}_{x(\tilde{p})}
  \left(\frac{\alpha_\p-2}{2},\frac{3-\alpha_\p}{2}\right)
  \right.\nonumber\\ 
&& +\quad 2\, \tilde{p}^{1-\alpha_\p}\left(\sqrt{1 + \tilde{p}^2} 
    - 1\right)\bigg]_{\tilde{p}_2}^{\tilde{p}_1}, \\
x(\tilde{p}) &=& (1 + \tilde{p}^2)^{-1}, \mbox{ and }
\tilde{p} = \frac{p_\p}{m_\p c},
\end{eqnarray}
where $\mathcal{B}_x(a,b)$ denotes the incomplete Beta-function
\citep{1965hmfw.book.....A} and $\tilde{p}_1$ and $\tilde{p}_2$ are the lower
and higher break momenta, respectively.

\subsubsection{Hadronically induced $\gamma$-ray emission}
\label{sec:gamma}

The CRp interact hadronically with the ambient thermal gas and produce pions,
provided their momentum exceeds the kinematic threshold $p_\rmn{thr} = 0.78
\mbox{ GeV }c^{-1}$ for the reaction.  The neutral pions decay into
$\gamma$-rays while the charged pions decay into secondary electrons (and
neutrinos):
\begin{eqnarray}
  \pi^\pm &\rightarrow& \mu^\pm + \nu_{\mu}/\bar{\nu}_{\mu} \rightarrow
  e^\pm + \nu_{e}/\bar{\nu}_{e} + \nu_{\mu} + \bar{\nu}_{\mu}\nonumber\\
  \pi^0 &\rightarrow& 2 \gamma \,.\nonumber
\end{eqnarray}
Only the CRp population above the kinematic threshold $p_\rmn{thr}$ is
visible through its decay products in $\gamma$-ray and synchrotron emission.

An analytic formula describing the omnidirectional (i.e. integrated over
$4\upi$ solid angle) differential $\gamma$-ray source function resulting from
$\pi^0$-decay of a power-law CRp population is given in \citet{2004A&A...413...17P}:
\begin{eqnarray}
\label{qgamma}
\lefteqn{
q_\gamma(\bmath{r},E_\gamma)\,\dd E_\gamma\,\dd V\simeq
\sigma_\rmn{pp}\,c\,n_\rmn{N}(\bmath{r})\,
2^{2-\alpha_\gamma}\,\frac{\tilde{n}_\crp(\bmath{r})}{\mbox{GeV}}}  \\
  & & \times\,\frac{4}{3\,\alpha_\gamma}\,\left( \frac{m_{\pi^0}\,c^2}
      {\mbox{GeV}}\right)^{-\alpha_\gamma}
      \left[\left(\frac{2\, E_\gamma}{m_{\pi^0}\, c^2}\right)^{\delta_\gamma} +
      \left(\frac{2\, E_\gamma}{m_{\pi^0}\, c^2}\right)^{-\delta_\gamma}
      \right]^{-\alpha_\gamma/\delta_\gamma}\!\dd E_\gamma\,\dd V, \nonumber
\end{eqnarray}
where $n_\rmn{N}(\bmath{r}) = d_\rmn{tar}\, n_\e(\bmath{r}) =
n_\e(\bmath{r})/(1 - \frac{1}{2} X_\rmn{He})$ denotes the target nucleon
density in the ICM while assuming primordial element composition with
$X_\rmn{He} = 0.24$, which holds approximately.  The formalism also includes
the detailed physical processes at the threshold of pion production like the
velocity distribution of CRp, momentum dependent inelastic CRp-p cross section,
and kaon decay channels.  The shape parameter $\delta_\gamma$ and the effective
cross section $\sigma_\rmn{pp}$ depend on the spectral index of the
$\gamma$-ray spectrum $\alpha_\gamma$ according to
\begin{eqnarray}
\label{delta}
\delta_\gamma &\simeq& 0.14 \,\alpha_\gamma^{-1.6} + 0.44\qquad\mbox{and} \\
\label{sigmapp}
\sigma_\rmn{pp} &\simeq& 32 \cdot
\left(0.96 + \rmn{e}^{4.4 \,-\, 2.4\,\alpha_\gamma}\right)\mbox{ mbarn}. 
\end{eqnarray}
There is a detailed discussion in \citet{2004A&A...413...17P} how the $\gamma$-ray
spectral index $\alpha_\gamma$ relates to the spectral index of the parent CRp
population $\alpha_\p$.  In Dermer's model, which is motivated by accelerator
experiments, the pion multiplicity is independent of energy yielding the
relation $\alpha_\gamma = \alpha_\p$ \citep{1986ApJ...307...47D,
  1986A&A...157..223D}. 

Provided the CRp population has a power-law spectrum, the integrated
$\gamma$-ray source density $\lambda_\gamma$ for pion decay induced
$\gamma$-rays can be obtained by integrating the $\gamma$-ray source function
$q_\gamma(\bmath{r}, E_\gamma)$ in Eq.~(\ref{qgamma}) over an energy interval
yielding
\begin{eqnarray}
\label{lambda_gamma}
\lambda_\gamma(\bmath{r}, E_1, E_2) &=& \int_{E_1}^{E_2} \dd E_\gamma\,
q_\gamma(\bmath{r}, E_\gamma) \\
\label{lambda_gamma2}
&=& A_\gamma(\alpha_\p)\, A_{E_\p}^{-1}(\alpha_\p)\, 
X_\crp(\bmath{r})\,n_\e^2(\bmath{r})\,k T_\e(\bmath{r}),\\
\mbox{where~~} 
A_\gamma(\alpha_\p) &=& 
\frac{\sigma_\rmn{pp}(\alpha_\p)\,c\,d_\e\,d_\rmn{tar}
      \left[\mathcal{B}_x\left(\frac{\alpha_\gamma + 1}{2\,\delta_\gamma},
      \frac{\alpha_\gamma - 1}{2\,\delta_\gamma}\right)\right]_{x_1}^{x_2}}
     {\left(\frac{m_{\pi^0}\,c^2}{\mbox{GeV}}\right)^{\alpha_\gamma-1}\,
      2^{\alpha_\gamma -1}\, \alpha_\gamma\,\delta_\gamma},\label{Agamma}\\
\mbox{and~~}
x_i &=& \left[1+\left(\frac{m_{\pi^0}\,c^2}{2\,E_i}
      \right)^{2\,\delta_\gamma}\right]^{-1} \mbox{~for~~}
      i \in \{1,2\}.
\end{eqnarray}
The $\gamma$-ray number flux $\mathcal{F}_\gamma$ is derived by means of volume
integration over the emission region and correct accounting for the growth of
the area of the emission sphere on which the photons are distributed:
\begin{equation}
  \label{eq:flux_gamma}
  \mathcal{F}_\gamma (E_1, E_2) = \frac{ 1+z}{4\upi \,D^2} \int \dd V\,
  \lambda_\gamma[\bmath{r}, (1+z) E_1, (1+z) E_2].
\end{equation}
Here $D$ denotes the luminosity distance and the additional factors of $1+z$
account for the cosmological redshift of the photons.

\subsubsection{Hadronically induced synchrotron emission}

Following for instance \citet{2000A&A...362..151D} and \citet{2004A&A...413...17P}, the
steady-state $\cre$ spectrum is governed by injection of secondaries and
cooling processes so that it can be described by the continuity equation
\begin{equation}
\frac{\upartial }{\upartial E_\e} \left( \dot{E_\e}(\bmath{r},E_\e) f_\e
(\bmath{r},E_\e) \right) = q_\e(\bmath{r}, E_\e)\,.
\end{equation}
For $\dot{E_\e}(\bmath{r},p) < 0$ this equation is solved by
\begin{equation}
f_\e (\bmath{r},E_\e) = \frac{1}{|\dot{E_\e}(\bmath{r},E_\e)|} \int_{E_\e}^\infty
\! \dd E_\e'  q_\e(\bmath{r}, E_\e')\,.
\end{equation}
For the energy range of interest, the cooling of the radio emitting CRe is
dominated by synchrotron and inverse Compton losses:
\begin{equation}
- \dot{E_\e}(\bmath{r},E_\e) = \frac{4\,\sigma_\mathrm{T}\, c}{3\,m_\e^2\,c^4} 
\left[\eps_B(\bmath{r}) + \eps_\rmn{CMB}\right]\,E_\e^2\,,   
\end{equation}
where $\sigma_\rmn{T}$ is the Thomson cross section, $\eps_B(\bmath{r})$ is the
local magnetic field energy density, and $\eps_\rmn{CMB} =
B^2_\rmn{CMB}/(8\upi)$ is the energy density of the cosmic microwave background
expressed by an equivalent field strength $B_{\rm CMB} = 3.24\,
(1+z)^2\umu\mbox{G}$.  Assuming that the parent CRp population is represented
by a power-law (\ref{fp}), the CRe population above a GeV is therefore
described by a power-law spectrum
\begin{eqnarray}
\label{fe_hadr}
f_\e (\bmath{r},E_\e) &=& \frac{\tilde{n}_\cre(\bmath{r})}{\rm GeV}
\,\left( \frac{E_\e}{\rm GeV} \right)^{-\alpha_\e}, \\
\label{nCRe}
\mbox{and~~}
\tilde{n}_\cre(\bmath{r}) &=& A_{\eps_\rmn{eff}}(\bmath{r})\,
\frac{\tilde{n}_\crp(\bmath{r})}{\eps_B(\bmath{r}) + \eps_\rmn{CMB}},\\
A_{\eps_\rmn{eff}} (\bmath{r}) &=& 
  \frac{16^{-(\alpha_\e-2)}}{\alpha_\e - 2}\,
  \frac{\sigma_\rmn{pp}\, m_\e^2\,c^4 \,n_\rmn{N}(\bmath{r})}
     {\sigma_\rmn{T} \,\rmn{GeV}} \,,
\end{eqnarray}
where the effective CRp-p cross section $\sigma_\rmn{pp}$ is given by
Eq.~(\ref{sigmapp}).

The hadronically induced synchrotron emissivity $j_\nu$ at frequency $\nu$ and
per steradian of such a CRe population (\ref{fe_hadr}) which is located in an
isotropic distribution of magnetic fields within the halo volume is given by
Eq.~(\ref{jnu}). However, the normalisation $\tilde{n}_\cre(\bmath{r})$ of the
CRe population is given by Eq.~(\ref{nCRe}). The spectral index of the
synchrotron emission is related to the CRp spectral index by $\alpha_\nu =
(\alpha_\e-1)/2 = \alpha_\p/2$.

% --- section: Minimum energy criterion --- %
\section{Minimum energy criteria}
\label{sec:MEC}

This section develops minimum energy criteria in order to estimate the magnetic
field and studies the tolerance region of the obtained estimates.  As described
in Sect.~\ref{sec:intro}, we discuss two different approaches when requiring
the non-thermal energy density of the source to be minimal for a particular
(observed) synchrotron emission. The classical minimum energy criterion used in
radio astronomy \citep{1993A&A...270...91P, 1996ARA&A..34..155B,
  1997A&A...325..898B} can be applied irrespective of the particular
acceleration process of CRe, but unfortunately it relies on uncertain
assumptions or parameters.  Provided the hadronic scenario of synchrotron
emission applies, these dependencies can be softened. The resulting minimum
energy argument needs to be changed accordingly.

The philosophy of these approaches is to provide an estimate of the
energetically least expensive radio emission model possible in each of these
physically motivated scenarios. Thus, the obtained minimum energy estimates
should not be taken literally in a sense that they are necessarily realised in
Nature. However, the minimum energy estimates allow scrutinising the hadronic
model for extended radio synchrotron emission in clusters of galaxies by
comparing to the thermal energy density.  For the classical minimum energy
estimate, such a comparison can yield important constraints on the accessible
parameter space.

The non-thermal energy density in the intra-cluster medium (ICM), which is the
quantity to be minimised, is composed of the sum of the energy densities in
magnetic fields, CRp and CRe:
\begin{equation}
  \label{eq:MEC:eps_NT}
  \eps_\NT = \eps_B + \eps_\crp + \eps_\cre.
\end{equation}
The CRp population also includes higher mass nuclei in addition to protons.
For convenience, we introduce canonical dimensionless energy densities by means
of scaling with the critical magnetic energy density $\eps_{B_\rmn{c}} =
B_\rmn{c}^2 / (8\upi)$, where $B_\rmn{c}$ is defined in Eq.~(\ref{Bc}):
\begin{eqnarray}
\label{eq:x_parameter}
\lefteqn{
x_\NT(\bmath{r})   \equiv \frac{\eps_\NT(\bmath{r})}{\eps_{B_\rmn{c}}}\,,
~x_\crp(\bmath{r}) \equiv \frac{\eps_\crp(\bmath{r})}{\eps_{B_\rmn{c}}}\,,
~x_\cre(\bmath{r}) \equiv \frac{\eps_\cre(\bmath{r})}{\eps_{B_\rmn{c}}}\,,}
\\
\lefteqn{
x_{B}(\bmath{r})   \equiv \frac{\eps_{B}(\bmath{r})}{\eps_{B_\rmn{c}}}\,,
~x_\rmn{th}(\bmath{r})\equiv \frac{\eps_\rmn{th}(\bmath{r})}{\eps_{B_\rmn{c}}}\,,
\mbox{ and }
x_\rmn{CMB} \equiv \frac{\eps_\rmn{CMB}}{\eps_{B_\rmn{c}}}\,.}
\end{eqnarray}
After presenting the conceptually simpler classical minimum energy criterion, we
will subsequently discuss the hadronic minimum energy criterion.

\subsection{Classical minimum energy criterion}

This section presents the classical minimum energy criterion from a physically
motivated point of view, probing synchrotron emission from CRe characterised by
a power-law distribution function without specifying their origin.  This
approach implies putting up with a dependence of the inferred magnetic field
strength on unknown parameters like the lower energy cutoff of the CRe
population or the unknown contribution of CRp to the non-thermal energy
density.

\subsubsection{Derivation}
Assuming a proportionality between the CRe and CRp energy densities,
i.e.~$x_\crp = k_\p x_\cre$, the non-thermal energy composition equation
(\ref{eq:MEC:eps_NT}) can be written as
\begin{equation}
  \label{eq:x_NT_class}
  x_\NT = x_B + (1 + k_\p) x_\cre.
\end{equation}
This assumption is reasonable if for instance the thermal electron population
and the CRp were energised by the same shock wave assuming that there is a
constant fraction of energy going into the CRp population by such an
acceleration process, provided injection processes alone determine the
energies.  In order to proceed, we need an expression for $j_\nu$ (given by
Eq.~(\ref{jnu})) as a function of the dimensionless energy densities considered
here:
\begin{eqnarray}
  \label{eq:jnu_class}
  j_\nu(\bmath{r}) &=&
  \frac{A_{E_\rmn{syn}}(\alpha_\e)\,\eps_{B_\rmn{c}}}{A_{E_\e}}\,
  x_\cre(\bmath{r})\,x_B(\bmath{r})^{1+\delta},\quad\mbox{where}\\
  \delta &=& \frac{\alpha_\nu - 1}{2} = \frac{\alpha_\e - 3}{4} .
\end{eqnarray}
For consistency reasons, which will become clear in Sect.~\ref{sec:hadronic}, we
introduce $\delta$ as a parametrisation of the spectral index. For typical
radio \mbox{(mini-)halos}, $\delta$ is a small quantity: The synchrotron
spectral index $\alpha_\nu = 1$ corresponds to $\delta = 0$.  Solving
Eq.~(\ref{eq:jnu_class}) for the CRe energy density yields
\begin{eqnarray}
  \label{eq:x_CRe_class}
  x_\cre &=& C_\rmn{class}(\bmath{r})\, x_B(\bmath{r})^{-1-\delta},
  \quad\mbox{where}\\
  C_\rmn{class}(\bmath{r}) &\equiv &
  \frac{A_{E_\e}\,j_\nu(\bmath{r})}
       {A_{E_\rmn{syn}}\,\eps_{B_\rmn{c}}}
  \propto \frac{j_\nu}{\nu^3}
\end{eqnarray}
is a convenient auxiliary variable.  Combining Eqs.~(\ref{eq:x_NT_class}) and
(\ref{eq:x_CRe_class}) yields the non-thermal energy density solely as a
function of the magnetic energy density
\begin{equation}
  \label{eq:x_NT_class2}
  x_\NT = x_B + (1 + k_\p) \,C_\rmn{class}(\bmath{r})\, 
  x_B(\bmath{r})^{-1-\delta}.
\end{equation}

Requesting this energy density to be minimal for a given synchrotron emissivity
yields the energetically least expensive radio emission model possible in this
approach:
\begin{equation}
\label{eq:energy_cond_class}
\left(\left.\frac{\upartial x_\NT}{\upartial x_B}\right|_{j_\nu}\right)
= 1 - (1 + k_\p) (1 + \delta)\, C_\rmn{class}(\bmath{r}) 
x_B(\bmath{r})^{-2-\delta}\stackrel{!}{=}0.
\end{equation}
The corresponding CRp, CRe and magnetic energy densities are given by
\begin{eqnarray}
  \label{eq:xB_min_class}
  x_{B_\rmn{min}}(\bmath{r}) &=& 
  \left[(1 + k_\p) (1 + \delta)\, 
    C_\rmn{class}(\bmath{r})\right]^{1/(2+\delta)}, \\
  x_{\cre_\rmn{min}}(\bmath{r}) &=& C_\rmn{class}(\bmath{r})
  \left[(1 + k_\p) (1 + \delta)\, 
    C_\rmn{class}(\bmath{r})\right]^{-(1+\delta)/(2+\delta)},\\
  x_{\crp_\rmn{min}}(\bmath{r}) &=& 
  k_\p\, x_{\cre_\rmn{min}}(\bmath{r}).
\end{eqnarray}
Note that these formulae deviate from \citet{1970ranp.book.....P}, since we use
a fixed interval in CRe energy rather than in radio frequency
(c.f.~Sect.~\ref{sec:intro}). However, they are equivalent to those obtained by
\citet{1997A&A...325..898B}.

\subsubsection{Localisation of classical minimum energy densities}

We wish to quantify how tight our statements about the inferred minimum energy
densities are, i.e.~to assign a tolerance region to the minimum energy
estimates. This region would have the meaning of a quasi-optimal realisation of
the particular energy densities.  The curvature radius at the extremal value is
one possible way of characterising the `sharpness' of the minimum:\footnote{One
  could picture this approach by assuming Gaussian statistics for the
  distribution of non-thermal energy densities. The curvature radius at the
  minimum would then correspond to the width $\sigma_B$ and thus yielding the
  $68\%$-confidence level with respect to this extremal value.}
\begin{equation}
  \label{eq:curvature_lin}
  \sigma_{x_B,\rmn{Gauss}} \equiv \left(\left. \frac{1}{x_\NT}
    \frac{\upartial^2 x_\NT}{\upartial x_B^2}\right|_{x_{B_\rmn{min}}}
  \right)^{-1/2}.
\end{equation}
In order to avoid unphysical negative values for the lower tolerance level of
$x_{B_\rmn{min}}$ we rather adopt the following logarithmic measure of
the curvature:
\begin{equation}
  \label{eq:sigma_B1}
  \sigma_{\ln x_B} \equiv \left(\left. 
    \frac{\upartial^2 \ln x_\NT}{\upartial (\ln x_B)^2}\right|_{x_{B_\rmn{min}}}
  \right)^{-1/2}.
\end{equation}
Considering the linear representation of $x_B$, this definition explicitly
implies tolerance levels which are given by $\exp(\ln x_B \pm \sigma_{\ln
  x_B})$.  Applying this definition to Eq.~(\ref{eq:x_NT_class2}) yields the
theoretical tolerance level of the estimated minimum magnetic energy
density,
\begin{equation}
  \label{eq:sigma_B2}
  \sigma_{\ln x_B}(\bmath{r}) = \left\{
    \frac{x_{B_\rmn{min}}^{-2-\delta}\,\left[C_\rmn{class}\, 
        \left(1 + k_\p\right) + x_{B_\rmn{min}}^{2+\delta}\right]^2}
         {C_\rmn{class} \left(1 + k_\p \right) 
           (2 + \delta)^2}\right\}^{1/2}(\bmath{r}).
\end{equation}
The tolerance level of the estimated minimum CRp energy density is given by 
\begin{equation}
  \label{eq:sigma_CRp1}
  \sigma_{\ln x_\crp} \equiv \left|\frac{\upartial \ln x_\crp}{\upartial \ln x_B}
  \right|_{x_{B_\rmn{min}}} \sigma_{\ln x_B},
\end{equation}
while the theoretical tolerance level of the estimated minimum CRe energy
density can be obtained likewise.  Applying this Gaussian error propagation to 
Eq.~(\ref{eq:x_CRe_class}), we obtain following general result:
\begin{equation}
  \label{eq:sigma_CRp2}
  \sigma_{\ln x_\crp} = \sigma_{\ln x_\cre} = (1+\delta)\,\sigma_{\ln x_B}.
\end{equation}

\subsubsection{Equipartition condition}
In order to investigate under which conditions the classical minimum energy
criterion implies exact equipartition, we examine the special case $\delta = 0$
of Eq.~(\ref{eq:xB_min_class}).  The resulting minimal magnetic and CRp energy
densities read
\begin{eqnarray}
  \label{eq:B_equipartition}
  x_{B_{0,\rmn{min}}}(\bmath{r}) &=& 
  \sqrt{(1+ k_\p)\, C_\rmn{class}(\bmath{r})}\,,\\
  x_{\crp_{0,\rmn{min}}}(\bmath{r}) &=& 
  \sqrt{\frac{k_\p^2\, C_\rmn{class}(\bmath{r})}{(1+ k_\p)}}
  \quad\stackrel{k_\p \gg 1}{\longrightarrow}\quad 
  x_{B_{0,\rmn{min}}}(\bmath{r})\,.
\end{eqnarray}
This comparison shows that there exists exact equipartition between the CRp and
magnetic energy densities if $\delta = 0$ and $k_\p \gg 1$!  In our Galaxy
$k_\p \simeq 100$ \citep{1996ARA&A..34..155B} and $\alpha_\nu=\alpha_\rmn{GHz}
= 0.8$ suggesting $\delta = -0.1$ \citep{1985A&A...153...17B}, which implies
that these equipartition conditions are well fulfilled.  Comparing this result
with studies that are using a combination of synchrotron emission, the local
CRe density, and diffuse continuum $\gamma$-rays, \citet{2000ApJ...537..763S}
interestingly imply the same magnetic field strength as inferred from
equipartition arguments \citep{1996ARA&A..34..155B}.  The corresponding minimal
CRe energy density is given by
\begin{equation}
  \label{eq:CRe_equipartition}
  x_{\cre_{0,\rmn{min}}}(\bmath{r}) \quad = \quad
  \sqrt{\frac{C_\rmn{class}(\bmath{r})}{(1+ k_\p)}}
  \quad\stackrel{k_\p \ll 1}{\longrightarrow}\quad 
  x_{B_{0,\rmn{min}}}(\bmath{r})\,.
\end{equation}
The classical minimum energy densities in CRe and magnetic fields are in exact
equipartition if $\delta = 0$ and $k_\p \ll 1$!  In the limiting case
$\delta=0$, the theoretical tolerance levels of the estimated classical
minimum energy densities read
\begin{equation}
  \label{eq:sigma_class}
  \sigma_{\ln x_{B_0}} = \sigma_{\ln x_{\crp_0}} = \sigma_{\ln x_{\cre_0}} = 1.
\end{equation}

\subsection{Hadronic minimum energy criterion}
\label{sec:hadronic}

This section deals with the hadronic minimum energy criterion and thus probes
hadronically induced synchrotron emission.  If a significant part of the CRe
population is known to be generated by hadronic interactions, the minimum
energy estimate allows testing for the realisation of the minimum energy state
for a given independent measurement of the magnetic field strength.  Thus, the
proposed minimum energy criterion would provide information about the dynamical
state of the magnetic field. 

\subsubsection{Derivation}

To derive the hadronic minimum energy criterion, we need an expression for the
CRp energy density as a function of the magnetic energy density which is
obtained by combining Eqs.~(\ref{jnu}) and (\ref{nCRe}):
\begin{eqnarray}
\label{eq:eps_CRp}
x_\crp(\bmath{r}) &=& C_\hadr(\bmath{r})\, [x_B(\bmath{r}) + 
x_\rmn{CMB}]\, x_B(\bmath{r})^{-1-\delta},\\
\label{eq:C_hadr}
C_\hadr(\bmath{r}) &\equiv&
  \frac{A_{E_\p}}{A_{E_\rmn{syn}}}
  \frac{j_\nu(\bmath{r})}{A_{\eps_\rmn{eff}}(\bmath{r})}
  \propto \frac{j_\nu(\bmath{r})}{\nu\,n_\rmn{N}(\bmath{r})},\\
\delta &=& \frac{\alpha_\nu - 1}{2} = \frac{\alpha_\e - 3}{4} 
        = \frac{\alpha_\p - 2}{4}.
\end{eqnarray}
The parameter $C_\hadr(\bmath{r})$ has the meaning of a hadronic synchrotron
emissivity per target nucleon density and per frequency. Its value decreases for
any existing cutoff in the parent CRp population as described by
Eq.~(\ref{cutoffs}).  For convenience we introduce the parameter $\delta$ which
ranges within $[-0.1,0.2]$ for conceivable CRp spectral indices $\alpha_\p =
[1.6,2.8]$.

Assuming the hadronic electron source to be dominant, we can neglect the primary
CRe population. In any case, the energy density of hadronically generated CRe
is negligible compared to the energy density of the parent CRp population:
Above energies of $\sim$~GeV the differential hadronic $\tilde{k}_\p =
\tilde{n}_\crp / \tilde{n}_\cre$ has typically values ranging between
$\tilde{k}_\p \sim 100$ (Perseus) and $\tilde{k}_\p \sim 300$ (Coma), where we
inserted the typical values of the central density and magnetic field strength.
The tilde in $\tilde{k}_\p$ indicates the slightly modified definition compared
to the previous classical case.  Requiring minimum non-thermal energy density
within the hadronic framework for a given synchrotron emissivity yields
\begin{equation}
\label{eq:energy_cond}
\left(\left.\frac{\upartial x_\NT}{\upartial x_B}\right|_{j_\nu}\right)
= 1 + \left.\frac{\upartial x_\crp}{\upartial x_B}
  \right|_{j_\nu}\stackrel{!}{=}0.
\end{equation}
By using Eq.~(\ref{eq:eps_CRp}) we obtain the following implicit equation for
the minimum magnetic energy density:
\begin{equation}
\label{eq:Bfield}
\left[(1 + \delta)\, x_\rmn{CMB} + \delta x_{B_\rmn{min}}
    (\bmath{r})\right]\,x_{B_\rmn{min}}(\bmath{r})^{-2-\delta}
    = C_\hadr^{-1}(\bmath{r}).
\end{equation}
The definition of $C_\hadr(\bmath{r})$ (\ref{eq:C_hadr}) reveals an implicit
dependence on the parametrised spectral index $\delta$. However, the right hand
side of the minimum energy criterion (\ref{eq:Bfield}) is uniquely determined
for a given spectral index and an observed synchrotron emissivity at a
particular frequency. Thus, the minimum energy density of the magnetic field
giving rise to an observed synchrotron emission in the hadronic model can
either be obtained by solving Eq.~(\ref{eq:Bfield}) numerically or applying the
asymptotic expansion which will be developed in Sect.~\ref{sec:asymptotic}.

\subsubsection{Asymptotic expansion for $\delta \neq 0$}
\label{sec:asymptotic}

\begin{figure}
\resizebox{\hsize}{!}{\includegraphics{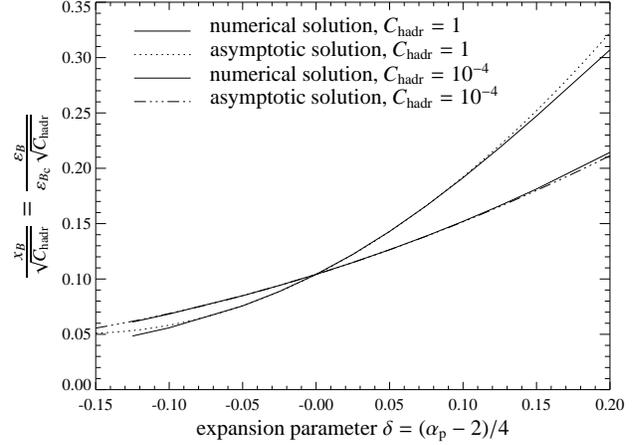}}
\caption{Comparison between the numerical solution and the second order asymptotic
  expansion for $C_\hadr=1$ (cool core cluster, formerly referred to as
    cooling flow cluster) and $C_\hadr=10^{-4}$ (cluster without a cool
    core).  The perfect match indicates a fast convergence of the asymptotic
  solution for the range of $\delta$ being considered.}
\label{fig:asymptotic}
\end{figure}

The asymptotic expansion of the magnetic field energy density as a function of
the small parameter $\delta$ follows from the minimum energy criterion
(\ref{eq:Bfield}):
\begin{eqnarray}
  \label{eq:asymptotic}
   x_{B_\rmn{min}}(\bmath{r}) &=& x_{B_0}(\bmath{r}) + 
     \delta x_{B_1}(\bmath{r}) + \delta^2 x_{B_2}(\bmath{r}) + O(\delta^3), \\ 
   x_{B_0}(\bmath{r}) &=& \sqrt{C_\hadr(\bmath{r})\, x_\rmn{CMB}}, \\ 
   x_{B_1}(\bmath{r}) &=& \frac{x_{B_0}}{2}\left(1 - \ln x_{B_0} +
     \frac{x_{B_0}}{x_\rmn{CMB}}\right)(\bmath{r}),\\
  \label{eq:asymptotic2}
   x_{B_2}(\bmath{r}) &=& \frac{x_{B_0}}{2}
   \left[-\frac{1}{2}\left(\ln x_{B_0}\right)^2
   -\frac{x_{B_1}}{x_{B_0}} \left(2 + \ln x_{B_0}\right) + 
   \frac{x_{B_1}^2}{x_{B_0}^2}\right].
\end{eqnarray}
A comparison between the numerical solution and the asymptotic solution for two
different values of the parameter $C_\hadr$ is shown in
Fig.~\ref{fig:asymptotic}.  The particular values of the parameter $C_\hadr$
will be motivated in Sect.~\ref{sec:appl_hadr}.  This figure should serve as
mathematical illustration of the convergence behaviour of the asymptotic
expansion while keeping the parameter $C_\hadr$ fixed.  The second order
asymptotic solution perfectly agrees with the exact solution for the range
$\delta = [-0.15,0.2]$ of conceivable CRp spectral indices $\alpha_\p =
[1.6,2.8]$.

\subsubsection{Localisation of hadronic minimum energy densities}
\label{sec:localization_hadr}

Considering the accuracy of the estimated minimum energy densities in the
hadronic scenario, we can also apply the logarithmic measure of the theoretical
tolerance level as defined in Eq.~(\ref{eq:sigma_B1}):
\begin{eqnarray}
  \label{eq:sigma_B_hadr}
  \lefteqn{\sigma_{\ln x_B}(\bmath{r}) \quad = \quad
    \left[x_{B_\rmn{min}}^{2+\delta} + C_\hadr
      \left(x_{B_\rmn{min}} + x_\rmn{CMB}\right)
    \right] (C_\hadr\, x_{B_\rmn{min}})^{-1/2}} \nonumber\\
\lefteqn{\,\,\times\left\{C_\hadr\, x_\rmn{CMB} +
    x_{B_\rmn{min}}^{1+\delta} 
    \left[(1+\delta)^2 x_{B_\rmn{min}} +
      (2+\delta)^2 x_\rmn{CMB}
    \right]
  \right\}^{-1/2}(\bmath{r}).}
\end{eqnarray}
The corresponding tolerance level for the minimum CRp energy density is
obtained by applying Gaussian error propagation (\ref{eq:sigma_CRp1}):
\begin{equation}
  \label{eq:sigma_CRp_hadr}
  \sigma_{\ln x_\crp}(\bmath{r}) = 
  \left(\frac{x_\rmn{CMB}}{x_{B_\rmn{min}}(\bmath{r}) + x_\rmn{CMB}} 
    + \delta\right)\, \sigma_{\ln x_B}(\bmath{r}).
\end{equation}
The consequences of these rather unwieldy formulae will become intuitively
clear in the next section where characteristic limiting cases are investigated.

\subsubsection{Special cases}
\label{sec:cases_hadr}
In order to gain insight into the hadronic minimum energy criterion, we
investigate special cases of Eq.~(\ref{eq:Bfield}), namely $\delta \to 0$,
$\eps_B \gg \eps_\rmn{CMB}$, and $\eps_B \ll \eps_\rmn{CMB}$, while
simultaneously considering the resulting tolerance regions of the previous
Sect.~\ref{sec:localization_hadr}.
\begin{enumerate}
\item $\delta \to 0$: The limit of $\delta \to 0$ corresponds to a hard
  spectral CRp population described by a spectral index of $\alpha_\p = 2.0$
  and thus $\alpha_\nu = 1$. There the dimensionless magnetic field energy
  density reads
  \begin{equation}
    \label{eq:x_B0}
    x_{B_{0,\rmn{min}}}(\bmath{r}) 
    = \sqrt{C_\hadr(\bmath{r})\, x_\rmn{CMB}} \propto 
    \left[\frac{j_\nu(\bmath{r})\,\eps_\rmn{CMB}}{\nu\,n_\rmn{N}(\bmath{r})}
      \right]^{1/2}.
  \end{equation}
  We can formulate the corresponding dimensionless CRp energy density resulting
  from this minimising argument:
  \begin{equation}
    \label{eq:xCRp0}
    x_{\crp_{0,\rmn{min}}}(\bmath{r}) = 
    C_\hadr(\bmath{r}) + \sqrt{C_\hadr(\bmath{r})\,
    x_\rmn{CMB}}.
  \end{equation}
  It is interesting to note that there are two regimes for
  $x_{\crp_{0,\rmn{min}}}$, namely
  \begin{eqnarray}
    \label{eq:xCRplimits}
    x_{\crp_{0,\rmn{min}}}(\bmath{r}) &=& 
    \left\{\begin{array}{l@{\quad\mbox{for }}l}
                C_\hadr(\bmath{r}), & 
                C_\hadr(\bmath{r})\gg 0.01\,\nu_\rmn{GHz}^{-2},\\
                x_{B_{0,\rmn{min}}}(\bmath{r}), &
                C_\hadr(\bmath{r})\ll 0.01\,\nu_\rmn{GHz}^{-2}\,.
            \end{array}
    \right.
  \end{eqnarray}
  Only in the limit of a small parameter $C_\hadr(\bmath{r})$ and $\delta = 0$,
  there is an exact equipartition between the hadronic minimum energy densities
  in CRp and magnetic fields!
  
  The tolerance regions of the estimated minimum energy densities are also more
  intuitive to understand in the limit $\delta=0$ compared to the general case
  laid down in Eq.~(\ref{eq:sigma_B_hadr}):
  \begin{eqnarray}
    \label{eq:sigmaB0}
    \sigma_{\ln x_{B_0}}(\bmath{r}) &=& 
      \left(1 + \frac{x_{B_{0,\rmn{min}}}(\bmath{r})}
        {2 x_\rmn{CMB}}\right)^{1/2},\quad\mbox{implying two regimes:}
      \nonumber\\
    \sigma_{\ln x_{B_0}}(\bmath{r}) &=& 
    \left\{\begin{array}{l@{\quad\mbox{for }}l}
                1, & x_{B_{0,\rmn{min}}} \ll x_\rmn{CMB},\\
                \left[\frac{\displaystyle x_{B_{0,\rmn{min}}}(\bmath{r})}
                     {\displaystyle 2 x_\rmn{CMB}}\right]^{1/2}, &
                x_{B_{0,\rmn{min}}} \gg x_\rmn{CMB}\,.
           \end{array}
    \right.
  \end{eqnarray}
  In the previously studied classical case (\ref{eq:sigma_class}), the
  tolerance region remains constant for all conceivable magnetic energy
  densities. In contrast to this, the hadronic scenario shows a increasing
  tolerance region $\sigma_{\ln x_{B_0}}$ for strong magnetic field strengths
  ($B_{0,\rmn{min}}\gg 3.24\, (1+z)^2~\umu\rmn{G}$) which is explained by the
  following argument: In the limit of strong magnetic fields, the hadronically
  induced synchrotron emission with $\alpha_\nu = 1$ does not depend any more
  on the magnetic field strength but only on the CRp energy density. This is,
  because the inverse Compton cooling is negligible in this regime, implying
  that observed synchrotron emission is insensitive to any variation of the
  magnetic field strength since all injected CRe energy results in synchrotron
  emission. Therefore magnetic field estimates inferred from minimum energy
  arguments are rather uncertain in the limit of strong magnetic field
  strengths.
  
  The tolerance levels of the corresponding CRp energy density, derived from
  Eq.~(\ref{eq:sigma_CRp_hadr}), shows two limiting regimes:
  \begin{eqnarray}
    \label{eq:sigma_CRp0}
     \sigma_{\ln x_{\crp_0}}(\bmath{r}) &=& 
      \left(1 + \frac{x_{B_{0,\rmn{min}}}(\bmath{r})}
        {x_\rmn{CMB}}\right)^{-1}
      \left(1 + \frac{x_{B_{0,\rmn{min}}}(\bmath{r})}
        {2 x_\rmn{CMB}}\right)^{1/2},\mbox{ implying}
      \nonumber\\
     \sigma_{\ln x_{\crp_0}}(\bmath{r}) &=& 
    \left\{\begin{array}{l@{\quad\mbox{for }}l}
                1, & x_{B_{0,\rmn{min}}} \ll x_\rmn{CMB},\\
                \left[\frac{\displaystyle x_\rmn{CMB}}
                           {\displaystyle x_{B_{0,\rmn{min}}}(\bmath{r})}
                \right]^{1/2}, &
                x_{B_{0,\rmn{min}}} \gg x_\rmn{CMB}\,.
           \end{array}
    \right.
  \end{eqnarray}
  The tolerance region of the inferred minimum CRp energy density decreases in the
  regime of strong magnetic fields which can be understood by the same token as
  above, i.e.~synchrotron losses dominate over the inverse Compton cooling in
  this limit. Thus, the observed radio emission reflects accurately the CRp
  energy density in the strong magnetic field limit.

\item $\eps_B \gg \eps_\rmn{CMB}$: In the limit of $\eps_B \gg
  \eps_\rmn{CMB}$, the magnetic field energy density is an even stronger
  function of the synchrotron emissivity and the number density of the ambient
  gas: 
  \alpheqn
  \begin{eqnarray}
    \label{eq:BggCMB}
    x_{B_\rmn{min}}(\bmath{r}) &=& \left[\delta\,C_\hadr(\bmath{r}) 
    \right]^{1/(1+\delta)} \\
    &\simeq& \delta\, C_\hadr(\bmath{r})^{1/(1+\delta)}
    \propto\delta\, \left[\frac{j_\nu(\bmath{r})}
        {\nu\,n_\rmn{N}(\bmath{r})}\right]^{1/(1+\delta)},
  \end{eqnarray}
  \reseteqn
  where we assumed $|\delta| \ll 1$ in the second step. In the limit of strong
  magnetic fields and small $\delta$, exact equipartition of the magnetic and
  CRp hadronic minimum energy density does not occur, because
  \begin{equation}
    \label{eq:epsCRp_BggCMB}
     x_{\crp_\rmn{min}}(\bmath{r}) = 
     \delta^{-\delta/(1+\delta)}C_\hadr(\bmath{r})^{1/(1+\delta)}
     \simeq C_\hadr(\bmath{r})^{1/(1+\delta)}(1 - \delta\,\ln\delta).
  \end{equation}

\item $\eps_\rmn{CMB} \gg \eps_B$: In the opposite limit, we obtain the
  following minimum energy criterion for the magnetic field energy density:
  \alpheqn
  \begin{eqnarray}
    \label{eq:CMBggB}
    x_{B_\rmn{min}}(\bmath{r}) &=& \left[(1 + \delta)\, x_\rmn{CMB}\, 
      C_\hadr(\bmath{r})\right]^{1/(2+\delta)} \\
    &\simeq& \left(1 + \frac{\delta}{2}\right)\,
    \left[x_\rmn{CMB}C_\hadr(\bmath{r})\right]^{1/(2+\delta)} \\
    &\propto& \left(1 + \frac{\delta}{2}\right)\,
    \left[\frac{j_\nu(\bmath{r})\eps_\rmn{CMB}}{\nu\,n_\rmn{N}(\bmath{r})}
    \right]^{1/(2+\delta)},
  \end{eqnarray}
  \reseteqn where we again assumed $\delta \ll 1$ in the second step.  In
  contrast to the previous limit, there is exact equipartition between the
  magnetic and CRp hadronic minimum energy density to zeroth order in $\delta$,
  \alpheqn
  \begin{eqnarray}
    \label{eq:epsCRp_CMBggB}
     x_{\crp_\rmn{min}}(\bmath{r}) &=& 
     (1+\delta)^{-(1+\delta)/(2+\delta)}
     \left[x_\rmn{CMB}C_\hadr(\bmath{r})\right]^{1/(2+\delta)} \\
     &\simeq& \left(1 - \frac{\delta}{2}\right)\,
    \left[x_\rmn{CMB}C_\hadr(\bmath{r})\right]^{1/(2+\delta)}.
  \end{eqnarray}
  \reseteqn
\end{enumerate}

% --- section: Future testing --- %
\section{Future testing}
\label{sec:testing}

This section will discuss possibilities of measuring the magnetic field
strength, averaged over the cluster volume, in order to test for the
realisation of the energetically least expensive state given by the minimum
energy criterion.

\subsection{Inverse Compton emission}

The CRe population seen in the radio band via synchrotron emission should also
scatter photons of the cosmic microwave background (CMB), the local radiation
field of elliptical galaxies, and the thermal X-ray emission of the ICM to
different energy bands \citep{1966ApJ...146..686F, 1967MNRAS.137..429R}.
Combining measurements of inverse Compton (IC) and synchrotron emission
eliminates the uncertainty in number density of the CRe population provided the
inevitable extrapolation of the CRe power-law distribution for certain observed
wavebands is justified.  This enables the determination of the magnetic field
strength $B$ for an IC detection and a lower limit on $B$ for a given
non-detection of the IC emission.

The source function $q_\rmn{IC}$ owing to IC scattering of CMB photons off
an isotropic power law distribution of CRe (Eq.~(\ref{fe})) is \citep[derived
from Eq.~(7.31) in][ in the case of Thomson scattering] {1979rpa..book.....R},
\begin{eqnarray}
\label{IC}
q_\rmn{IC}(\bmath{r},E_\gamma) &=& \tilde{q}(\bmath{r})\,
f_\rmn{IC} (\alpha_e)\, 
\left(\frac{m_\e\, c^2}{\mbox{GeV}}\right)^{1 - \alpha_\e}
\left(\frac{E_\gamma}{k T_\rmn{CMB}}\right)^{-(\alpha_\nu+1)}, \\
f_\rmn{IC} (\alpha_e) &=& \frac{2^{\alpha_\e+3}\, 
  (\alpha_\e^2 + 4\, \alpha_\e + 11)}
  {(\alpha_\e + 3)^2\, (\alpha_e + 5)\, (\alpha_e + 1)} \nonumber\\
&&\times\,\,\Gamma\left(\frac{\alpha_e + 5}{2}\right)\,
  \zeta\left(\frac{\alpha_e + 5}{2}\right)\,,\\
\mbox{and }\tilde{q}(\bmath{r}) &=& 
\frac{3\upi\, \sigma_\rmn{T}\, \tilde{n}_\cre(\bmath{r})\,
      \left(k T_\rmn{CMB}\right)^2\,}{h^3\, c^2}\,, 
\end{eqnarray}
where $\alpha_\nu = (\alpha_\e-1)/2$ denotes the spectral index, $\zeta(a)$ the
Riemann $\zeta$-function \citep{1965hmfw.book.....A}, and
$\tilde{n}_\cre(\bmath{r})$ is given by Eq.~(\ref{Xcrescaling}). After
integrating over the considered energy interval and the IC emitting volume in
the cluster, the particle flux $\mathcal{F}_\gamma(E_1, E_2)$ is obtained
(c.f.~Eq.~(\ref{eq:flux_gamma})).

\citet{1998AA...330...90E} compiled non-detection limits of IC emission of
different photon fields in various wavebands from the Coma cluster and obtained
the tightest limits on $B$ from the CMB photon field. The same CRe population
emitting radio synchrotron radiation scatters CMB photons into the hard X-ray
band.  Non-detection of this IC emission by the OSSE experiment
\citep{1994ApJ...429..554R} yields a lower limit on the central magnetic field
strength of $B_\rmn{Coma}(0) > 0.2~\umu\rmn{G}~f_B^{-0.43}$, where $f_B$ is the
filling factor of the magnetic field in the volume occupied by CRe.  Provided
the CRe power-law distribution can be extrapolated to lower energies, the limit
given by the EUV flux \citep{1996Sci...274.1335L} predicts a magnetic field
strength stronger than $B_\rmn{Coma}(0) > 1.2~\umu\rmn{G}~f_B^{-0.43}$.  

The reported high energy X-ray excess of the Coma cluster by the Beppo-Sax
satellite \citep{1999ApJ...513L..21F} initiated other theoretical explanations
about the origin of such an excess than IC up-scattering of CMB photons by
relativistic electrons. One possibility implies the existence of a
bremsstrahlung emitting supra-thermal electron population between 10 and
100~keV which would also produce a unique Sunyaev-Zel'dovich signature
\citep{1999AA...344..409E, 2000A&A...360..417E, 2000ApJ...535L..71B,
  2000ApJ...532L...9B, 2002MNRAS.337..567L, 2003A&A...397...27C}. However, such
a population is questioned on theoretical reasons \citep{2001ApJ...557..560P},
and even the high X-ray excess of Coma itself is under debate
\citep{2004A&A...414L..41R, 2004ApJ...602L..73F}.  The data analysis of the RXTE
observation of A~2256 yielded also evidence for a second spectral component
\citep{2003ApJ...595..137R}.  On the basis of statistics alone, the detected
emission is inconclusive as to whether it originates from a thermal
multi-temperature fluid or an isothermal gas in combination with a non-thermal
IC power-law emission.  Future measurements with the IBIS instrument on-board
the INTEGRAL satellite should provide even tighter upper limits respectively
detections of the IC X-ray flux of a particular cluster and should therefore
allow even tighter lower limits on the magnetic field strength.

\subsection{$\bgamma$-ray emission}
\label{sec:gamma-testing}

This subsection outlines the method for estimating upper limits on the magnetic
field strength using hadronic CRp interactions. The method is based on the idea
of combining hadronically induced $\gamma$-ray and synchrotron emission to
eliminate the uncertainty in number density of the CRp population.  For this
purpose, one necessarily needs to resolve the detailed broad spectral signature
of $\gamma$-rays resulting from the $\pi^0$-decay ($\pi^0$-bump centred on
$m_{\pi^0} c^2/2 \simeq 67.5 \mbox{ MeV}$) as laid down in
Eq.~(\ref{lambda_gamma2}). This is to exclude other possible processes
contributing to diffuse extended $\gamma$-ray emission like IC radiation or
dark matter annihilation.  Because of possible other additional contributions
to the diffuse synchrotron emission from CRe populations, e.g.~primarily
accelerated electrons, we are only able to provide an upper limit on the
magnetic field strength.

The proposed algorithm allows for different spatial resolutions of the
$\gamma$-ray and synchrotron emission. The application, we have in mind, is the
determination of intracluster magnetic fields. In this case, $\gamma$-ray
observations of the $\pi^0$-decay induced $\gamma$-ray emission signature
are only able to provide integrated $\gamma$-ray fluxes of the entire cluster
due to their comparably large point spread function.  $\gamma$-ray fluxes
depend on the thermal electron density and temperature profiles which have to
be derived from X-ray observations.  However, if we assumed a comparable
resolution in $\gamma$-ray and synchrotron emission the dependences on the
thermal electron population could be eliminated (c.f.~Sect.~\ref{sec:CRp}).
The algorithm consists of the following two steps:
\begin{enumerate}
\item Choosing a constant scaling parameter $X_\crp$ for the CRp population and
  performing the volume integral of the energy integrated $\gamma$-ray source
  density $\lambda_\gamma$ (\ref{lambda_gamma2}) yields the $\gamma$-ray flux
  according to Eq.~(\ref{eq:flux_gamma}).  The CRp parameter $X_\crp$ is
  obtained by comparing the observed to the theoretically expected $\gamma$-ray
  flux.
\item Inserting $X_\crp$ into the synchrotron emissivity $j_\nu$ (\ref{jnu})
  enables us to solve for the magnetic field strength as function of angle on
  the sky when comparing to radio surface brightness observations.
\end{enumerate}

Once a detailed angular distribution of $\pi^0$-decay induced $\gamma$-ray
emission from a particular astrophysical object is available this algorithm may
be implemented for the average of pixels contained within a certain solid
angle. In this case the spatial distribution of CRp may even be deprojected.

Is there a chance to apply this method to galaxy cluster magnetic fields with
future $\gamma$-ray instruments?  Because of the necessity of resolving the
broad spectral signature of $\gamma$-rays resulting from the $\pi^0$-decay
centred on $\sim 67.5 \mbox{ MeV}$, the imaging atmospheric \v{C}erenkov
technique with a lower energy cutoff above 10~GeV is not applicable.
Contrarily, the LAT instrument on-board GLAST scheduled to be launched in 2007
has an angular resolution better than $3.5\degr$ at 100~MeV while covering an
energy range from 20~MeV up to 300~GeV with an energy resolution better than
10\%. Assuming a photon spectral index of $\alpha_\gamma=2$ for the
$\gamma$-ray background, the point-source sensitivity at high galactic latitude
in a one year all-sky survey is better than $6\times 10^{-9} \mbox{ cm}^{-2}
\mbox{ s}^{-1}$ for energies integrated above 100~MeV. Specifically, assuming a
CRp spectral index $\alpha_\p=2.3$ and a flat $X_\crp$ for simplicity, such a
one year all-sky survey is able to constrain $X_\crp < 0.01$ (Perseus) and
$X_\crp < 0.04$ (Coma). Taking additionally into account the $\gamma$-ray flux
between 20~MeV and 100~MeV as well as a longer survey time will improve the
sensitivities and yield even tighter limits on $X_\crp$.  Comparing these
limits with energetically favoured values of $X_{\crp_\rmn{min}}$ which are
obtained by applying hadronic minimum energy arguments to a given radio
synchrotron emission (c.f.{\ }Fig.~\ref{fig:MEC}) yields comparable values in
the case of Perseus while the situation in Coma is less optimistic. However, a
definitive answer to the applicability of this method can not be given on the
basis of minimum energy arguments because such a minimum energy state is not
necessarily realised in Nature.

% --- section: Application --- %
\section{Applications}
\label{sec:applications}
In this section we apply the classical and hadronic minimum energy criterion to
the radio (mini-)halos of the Coma and Perseus galaxy clusters. For simplicity,
the CRp and CRe spectral indices are assumed to be independent of position and
therefore constant over the cluster volume. If a radial spectral steepening as
reported by \citet{1993ApJ...406..399G} in the case of the Coma radio halo will
be confirmed by future radio observations evincing a better signal-to-noise
ratio, the CRp and CRe spectral index distributions would have to embody this
additional degree of freedom. We discuss in Sect.~\ref{sec:magComa} that a
moderate steepening would not significantly modify the hadronic minimum energy
condition while a strong steepening would challenge the hadronic scenario.

\subsection{Classical minimum energy criterion}
\label{sec:appl_class}

\begin{table}
\caption[t]{Individual parameters describing the extended diffuse radio
  emission in the Coma and Perseus galaxy cluster according to
  Eq.~(\ref{beta}). The maximal radius to which these profiles are applicable
  is denoted by $r_\rmn{max}$. The radio data at 1.4~GHz are taken from
  \citet{1997A&A...321...55D} (Coma) and \citet{1990MNRAS.246..477P}
  (Perseus) while the profile of the Coma cluster at 326~MHz is taken from 
  \citet{2001A&A...369..441G} which is based on radio observations by
  \citet{1990AJ.....99.1381V}.}   
\vspace{-0.1 cm}
\begin{center}
\begin{tabular}{cccrccc}
\hline
        & $S_0$ & \multicolumn{3}{c}{$r_\rmn{c}$} & $r_\rmn{max}$ & \\
Cluster & $[\mbox{Jy arcmin}^{-2}]$
        & \multicolumn{3}{c}{$[h_{70}^{-1}\mbox{ kpc}]$}
        & $[h_{70}^{-1}\mbox{ Mpc}]$ & $\beta$ \\
\hline 
\multicolumn{7}{l}{\underline{1.4~GHz observations:}} \\
\vphantom{\Large A}%
A1656 (Coma)    & $1.1 \times 10^{-3}$ && 450 && 1.0 & 0.78 \\
A426 (Perseus)  & $2.3 \times 10^{-1}$ && 30  && 0.1 & 0.55 \\
\hline 
\multicolumn{7}{l}{\underline{326~MHz observation:}} \\
\vphantom{\Large A}%
A1656 (Coma)    & $4.7 \times 10^{-3}$ && 850 && 0.7 & 1.07 \\
\hline 
\end{tabular}
\end{center}
\label{tab:radioprofiles}
\end{table}

\begin{figure*}
\begin{tabular}{cc}
\resizebox{0.48\hsize}{!}{\includegraphics{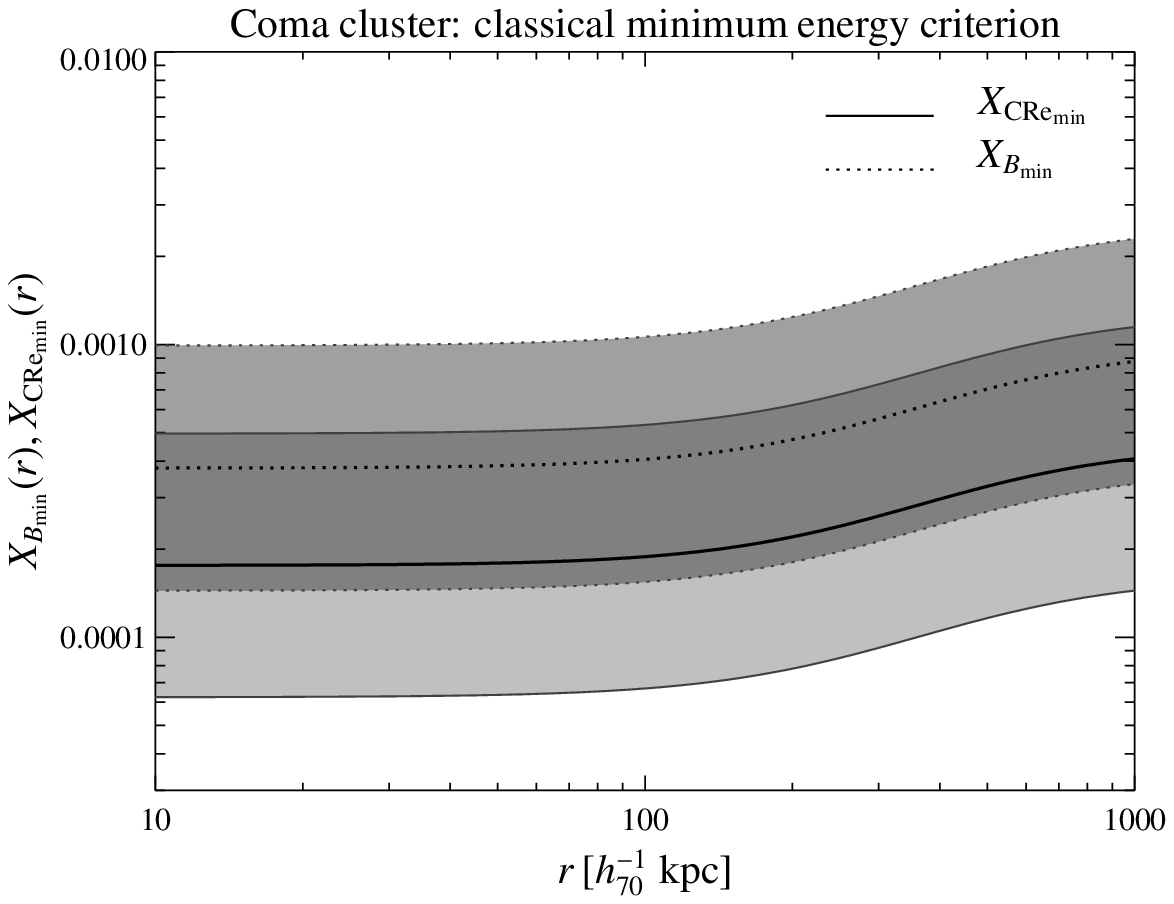}} &
\resizebox{0.48\hsize}{!}{\includegraphics{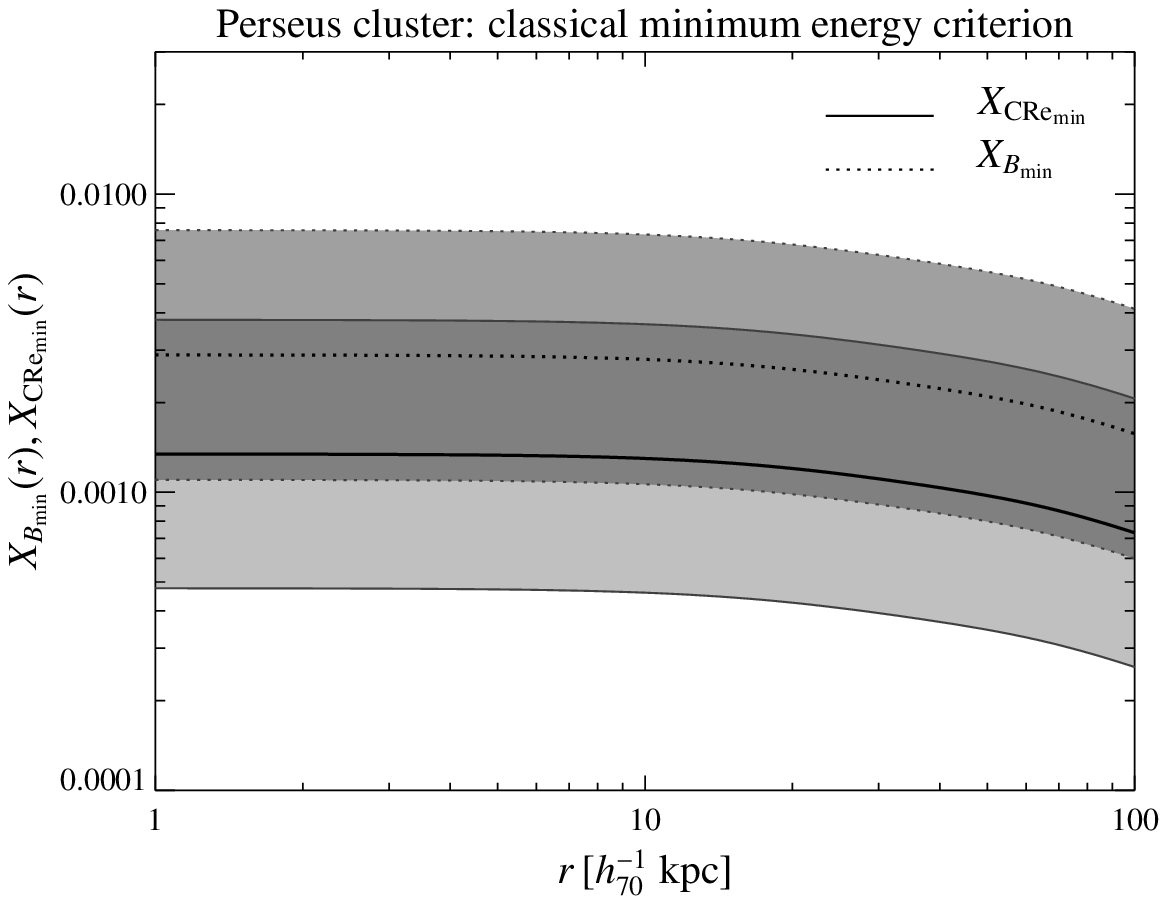}} \\
\\
\resizebox{0.48\hsize}{!}{\includegraphics{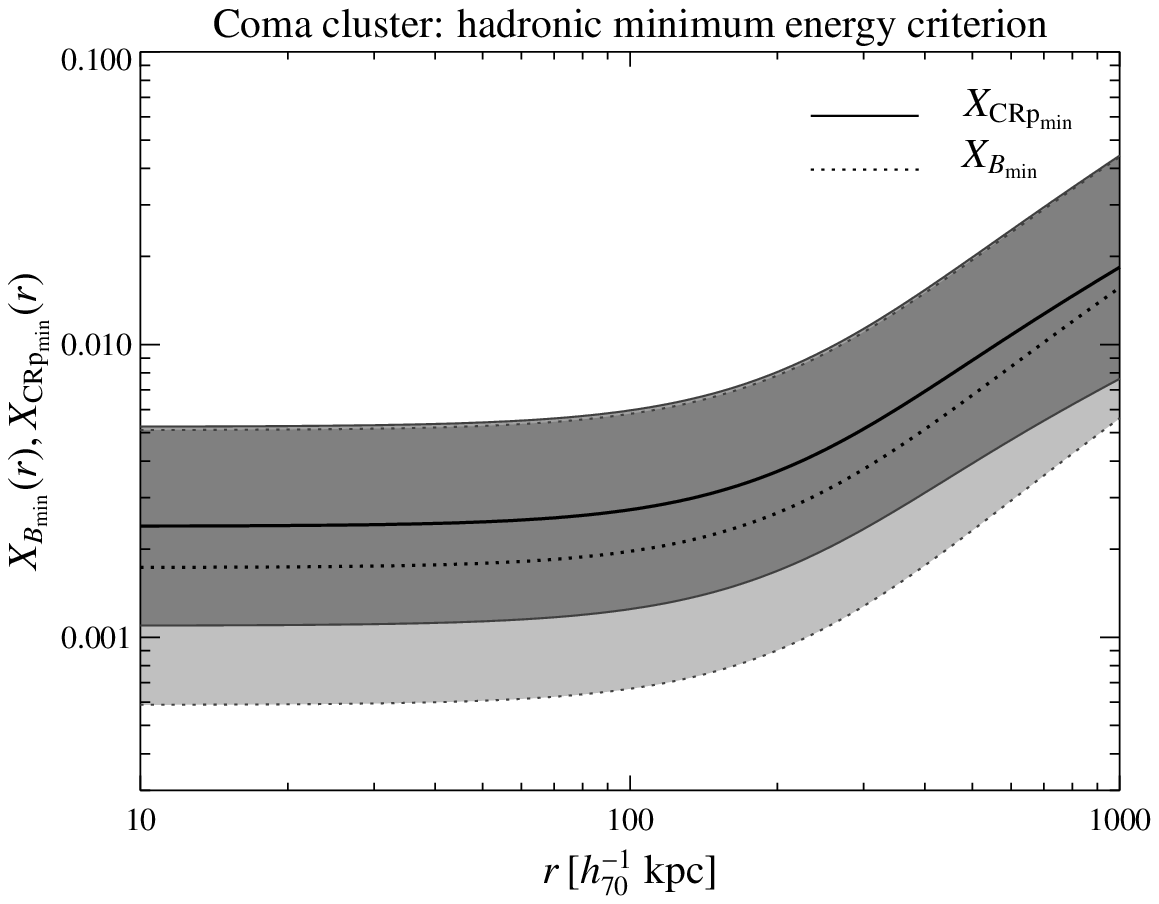}} &
\resizebox{0.48\hsize}{!}{\includegraphics{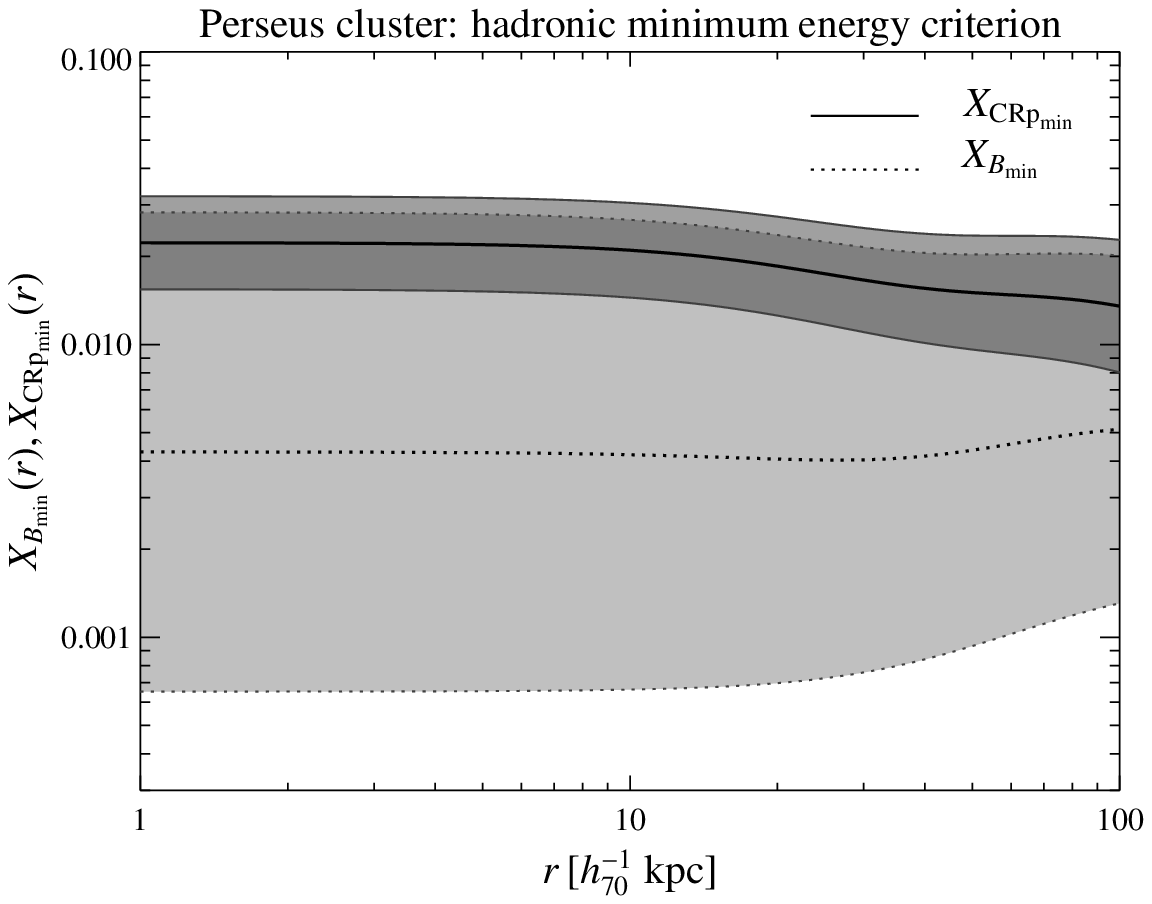}} 
\end{tabular}
\caption{Profiles of the CRe-to-thermal energy density
  $X_{\cre_\rmn{min}}(r)$ (solid) and magnetic-to-thermal energy density
  $X_{B_\rmn{min}}(r)$ (dotted) as a function of deprojected radius are shown.
  The different energy densities are obtained by means of the classical minimum
  energy criterion (upper panels) and the hadronic minimum energy criterion
  (lower panels). In the latter scenario, profiles of the scaled CRp energy
  density are shown instead of CRe profiles.  The left hand side shows profiles
  of the Coma cluster while the right hand side represents profiles of the
  Perseus cluster.  The light shaded areas represent the logarithmic tolerance
  regions of $X_{B_\rmn{min}}(r)$ and $X_{\cre_\rmn{min}}(r)$, respectively,
  while the dark shaded regions indicate the overlap and thus the possible
  equipartition regions in the quasi-optimal case.}
\label{fig:MEC}
\end{figure*}

In a first step we have to deproject the radio surface brightness and electron
density profiles: In analogy to X-ray observations we assume the azimuthally
averaged radio profile to be described by a $\beta$-model,
\begin{equation}
S_\nu (r_\bot)= S_0\, 
\left[ 1 + \left( \frac{r_\bot}{r_{\rmn{c}}}\right)^2\right]
^{-3\beta + 1/2}.
\label{beta}
\end{equation}
Deprojecting this profile yields the radio emissivity \citep[c.f.~Appendix
of][]{2004A&A...413...17P}
\begin{equation}
  \label{eq:Coma:radio}
  j_\nu (r) = \frac{S_0}{2\upi\, r_\rmn{c}}\, 
  \frac{6\beta - 1}{\left(1 + r^2/r_\rmn{c}^2\right)^{3 \beta}}\,
  \mathcal{B}\left(\frac{1}{2}, 3\beta\right)
  = j_{\nu,0} \left(1 + r^2/r_\rmn{c}^2\right)^{-3 \beta}.
\end{equation}
The individual parameters for the Coma radio halo and Perseus radio mini-halo
are shown in Table \ref{tab:radioprofiles}. Both profiles describe the extended
diffuse emission where all point sources have been subtracted.  Particularly,
the extremely bright flat-spectrum core owing to relativistic outflows of the
radio galaxy NGC~1275 in the centre of the Perseus cluster has been excluded
from the fit.  The electron density profiles are inferred from X-ray
observations by \citet{1992A&A...259L..31B} (Coma) and
\citet{2003ApJ...590..225C} (Perseus).

In the following, we assume that the energy distribution of the CRe population
above a MeV is represented by a power-law in energy $E_\e$ with a lower cutoff.
As a word of caution, such an assumption might be a strong simplification in
the case of turbulent acceleration models yielding more complex energy
distributions which are considerably flatter at lower energies
\citep{2001MNRAS.320..365B,2001ApJ...557..560P,2002ApJ...577..658O} . We assume
a CRe spectral index of $\alpha_\e = 3.3$ which translates into a synchrotron
spectral index of $\alpha_\nu = 1.15$. This is consistent with radio data of
Perseus and Coma; particularly when considering the spectral cutoff between 1
and 10 GHz owing to the Sunyaev-Zel'dovich flux decrement
\citep{1997A&A...321...55D, 2002A&A...396L..17E, 2004A&A...413...17P}.

Applying the classical minimum energy criterion to the diffuse synchrotron
emission of the Coma cluster yields a central magnetic field strength of
$B_\rmn{Coma}(r=0) = 1.1^{+ 0.7}_{- 0.4}~\umu\rmn{G}$. In the case of the
Perseus radio mini-halo we obtain $B_\rmn{Perseus}(r=0) =
7.2^{+4.5}_{-2.8}~\umu\rmn{G}$.  The indicated tolerance levels derive from
the logarithmic definition of the theoretical accuracies of the minimum in
Eq.~(\ref{eq:sigma_B1}). However, these values are highly dependent on the
lower energy cutoff of the CRe population, $E_1$, and the CRp proportionality
parameter $k_\p$. Following the philosophy of our paper we adopt a physically
motivated lower Coulomb cutoff of the CRe distribution of $E_1 = 0.1$~GeV
corresponding to a relativistic $\gamma$ factor of $\gamma_\e \simeq 200$
\citep[as suggested by][]{1999ApJ...520..529S}.  We also adopt a conservative
choice of the proportionality constant between the CRe and CRp energy densities
of $k_\p = 1$. An increase of $k_\p$ would directly increase the magnetic field
strength by approximately the square root of this factor.

For a cluster-wide comparison of energy densities of magnetic fields, CRe, and
CRp, it is convenient to introduce a scaling with the thermal energy density by
means of
\begin{equation}
  \label{eq:XB}
  X_B(\bmath{r}) \equiv \frac{x_B(\bmath{r})}{x_\rmn{th}(\bmath{r})} 
  = \frac{\eps_B(\bmath{r})}{\eps_\rmn{th}(\bmath{r})}, \mbox{ and}
\end{equation}
\begin{equation}
  \label{eq:XCRp}
  X_{\crp,\cre}(\bmath{r}) \equiv
  \frac{x_{\crp,\cre}(\bmath{r})}{x_\rmn{th}(\bmath{r})}  
  = \frac{\eps_{\crp,\cre}(\bmath{r})}{\eps_\rmn{th}(\bmath{r})}.
\end{equation}
Profiles of the CRe-to-thermal energy density $X_{\cre_\rmn{min}}(r)$ and
magnetic-to-thermal energy density $X_{B_\rmn{min}}(r)$ are shown in
Fig.~\ref{fig:MEC} for the Coma and Perseus cluster and our different
scenarios.  The upper panels show the scaled energy densities in the
acceleration model of CRe, obtained by the classical minimum energy criterion.
While the optimal magnetic energy density is roughly a factor of two larger
than the CRe energy density, they both can be in equipartition for the
quasi-optimal case of their distribution of energy densities, as indicated by
the dark shaded regions. In order to explain the observed synchrotron emission
in the CRe acceleration scenario the CRe and magnetic energy densities are only
required to be below one percent of the thermal energy density.

\subsection{Hadronic minimum energy criterion}
\label{sec:appl_hadr}

Assuming a CRp spectral index of $\alpha_\p = 2.3$ when applying the hadronic
minimum energy criterion to the diffuse synchrotron emission of the Coma
cluster yields a central magnetic field strength of $B_\rmn{Coma}(r=0) =
2.4^{+1.7}_{-1.0}~\umu\rmn{G}$. In the case of the Perseus cluster we obtain
$B_\rmn{Perseus}(r=0) = 8.8^{+13.8}_{-5.4}~\umu\rmn{G}$.  Both inferred
profiles of the magnetic field are relatively flat: While the magnetic field
strength in the outer part of the radio mini-halo in Perseus ($r\simeq
100~h_{70}^{-1} \mbox{ kpc}$) declines to a value of 55\% of its central
value, the magnetic field in the outer region of the radio halo in Coma
($r\simeq 1~h_{70}^{-1} \mbox{ Mpc}$) only decreases to 72\% of its central
value.

As discussed in Sect.~\ref{sec:cases_hadr}, the hadronic scenario shows an
increasing tolerance region for strong magnetic field strengths which are
expected to be present in the case of cool core clusters, such as Perseus
\citep{1993ApJ...416..554T,2002ARA&A..40..319C,2003A&A...412..373V}. In this
limit, synchrotron losses dominate over inverse Compton cooling. This almost
cancels the dependence of the synchrotron emissivity on the magnetic energy
density.  The lower panels of Fig.~\ref{fig:MEC} show the scaled energy
densities $X_{B_\rmn{min}}(r)$ and $X_{\crp_\rmn{min}}(r)$ as inferred from the
hadronic scenario.  In both clusters the optimal CRp energy density is larger
than the magnetic energy density within this model. However, both energy
densities can be again in equipartition for the quasi-optimal case of their
distribution, as indicated by the dark shaded regions.

Owing to the inverse dependence of $\sigma_{\ln x_B}$ and $\sigma_{\ln x_\crp}$
on the magnetic energy density in the limit of strong magnetic fields
(c.f.~Eqs.~(\ref{eq:sigmaB0}) and (\ref{eq:sigma_CRp0})), a large tolerance
region of $X_{B_\rmn{min}}$ immediately implies a well defined localisation of
$X_{\crp_\rmn{min}}$.  In the Perseus cluster this results in a confinement for
the CRp energy density of $2\%\pm 1\%$ of the thermal energy density.  On the
other hand, in the Coma cluster $X_{B_\rmn{min}}(r)$ and
$X_{\crp_\rmn{min}}(r)$ are required to increase by less than one order of
magnitude from the centre to the outer parts of the cluster in order to account
for the observed radio halo.  This increase might be partly due to azimuthally
averaging the aspheric electron density distribution of the Coma cluster
\citep{2004A&A...413...17P}.

High values for the radio emissivity per target density and frequency $C_\hadr$
of order unity seem to reflect conditions in cool core clusters (formerly
referred to as cooling flow cluster) whereas smaller values seem to represent
conditions in clusters without cool cores as shown in the following:
\begin{equation}
  \label{eq:Cappl}
  C_\hadr \equiv \frac{A_{E_\p}}{A_{E_\rmn{syn}}}
  \frac{j_\nu}{A_{\eps_\rmn{eff}}} 
  = C_\rmn{cluster}\, \left(\frac{\nu}{1.4\mbox{ GHz}}\right)^{-1}
  \left(\frac{n_\e}{n_{\e,0}}\right)^{-1}
  \left(\frac{j_\nu}{j_{\nu,0}}\right),
\end{equation}
where $C_\rmn{Coma} = 9.4 \times 10^{-4}$ and $C_\rmn{Perseus} = 1.5
\times 10^{-1}$.

\subsection{Possibility of a hadronic scenario in Perseus and Coma}
\label{sec:magComa}

\subsubsection{Perseus radio mini-halo}
The azimuthally averaged radio surface brightness profile of the Perseus
mini-halo matches the expected emission by the hadronic scenario well on all
radii \citep{2004A&A...413...17P} while requiring almost flat profiles for CRp
and magnetic energy densities relative to the thermal energy density, $X_\crp$
and $X_B$, respectively.  Moreover, the small amount of required energy density
in cosmic ray protons $\eps_\crp$ ($\sim 2\%$ relative to the thermal energy
density) supports the hypothesis of a hadronic origin of the Perseus radio
mini-halo not only because the hadronic minimum energy criterion predicts a
close confinement of $\eps_\crp$ (see Sect.~\ref{sec:appl_hadr}) but also
because cosmological simulations carried out by \citet{2001ApJ...559...59M}
easily predict a CRp population at the clusters centre of this order of
magnitude.

\subsubsection{Coma radio halo}
The energetically favoured radial profile for the magnetic field strength in
the Coma cluster is almost flat as predicted by the hadronic minimum energy
criterion (see Sect.~\ref{sec:appl_hadr}).  Provided these results would be
realised in Nature, this apparently contradicts profiles of the magnetic field
strength as inferred from numerical simulations which seem to follow the
electron density $n_\e(r)$ according to $B(r) \propto n_\e(r)^{\alpha_B}$ with
$\alpha_B \in [0.5,0.9]$ \citep{1999A&A...348..351D, 2001A&A...378..777D}. It
would also contradict theoretical considerations assuming the magnetic field to
be frozen into the flow and isotropised, i.e.~$\alpha_B = 2/3$
\citep{1993MNRAS.263...31T}.  Applying the flux freezing conditions to the
electron density profile of Coma \citep{1992A&A...259L..31B} yields an expected
decline of the magnetic field strength from its central value to the magnitude
at $1\,h_{70}^{-1} \mbox{ Mpc}$ by a factor of $\sim 6.7$.
  
However, there are other numerical, physical, and observational arguments
indicating large uncertainties in the origin, amplification mechanism, and
specific profile of the magnetic field strength, thus leaving the hadronic
scenario as a viable explanation of the Coma radio halo: In contrast to the
cited numerical simulations, there are other cosmological simulations
\citep{2001CoPhC.141...17M, 2001ApJ...562..233M} which are able to produce
giant radio halos in the hadronic scenario and therefore reasonably flat
profiles of the magnetic field strength. From the physical point of view, there
could be stronger shear flows or a larger number of weaker shocks in the outer
parts of clusters which are unresolved or not accounted for in current
simulations.  This would imply stronger additional amplification of the
magnetic field strength in the outer parts of clusters yielding a flatter
profile of the magnetic field strength.  Observationally, there are still
uncertainties in the radio surface profiles which are increased by azimuthally
averaging the diffuse synchrotron emission in the presence of non-centrally
symmetric emission components such as the so-called radio-bridge in Coma around
NGC~4839.

\begin{figure}
\resizebox{\hsize}{!}{\includegraphics{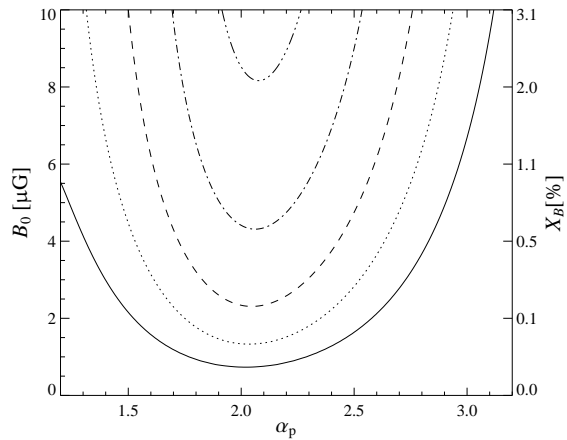}}
\caption{Parameter study on the ability of hadronically originating CRe to
  account for the radio halo of Coma. Assuming the profile of the magnetic
  field to scale with the square root of the electron density yields a flat
  magnetic-to-thermal energy density ratio $X_B$. Shown are contour lines from
  the bottom to the top of $\mbox{max}(X_\crp) = (1,0.3,0.1,0.03,0.01)$ for the
  range $r\le 1\,h_{70}^{-1}\mbox{ Mpc}$ in parameter space spanned by
  $\alpha_\p$ and $B_0$.  Conservative choices for CRp spectral breaks have
  been assumed.  The lower part represents the region in parameter space, where
  the hadronic scenario faces serious challenges for explaining the observed
  radio halo of Coma.}
\label{fig:parameter}
\end{figure}

In order to account for the radio halo of Coma in the hadronic scenario, the
product of $X_\crp$ and $X_B$ needs to increase by nearly two orders of
magnitude towards the outskirts of the halo (c.f.{\ }lower left panel of
Fig.~\ref{fig:MEC}). Leaving aside the minimum energy criterion, this increase
can be split arbitrarily among the magnetic and CRp energy density ratios.  For
instance a constant magnetic-to-thermal energy density ratio $X_B$,
corresponding to $\alpha_B = 0.5$ in an isothermal cluster, is still consistent
within the theoretically expected tolerance regions, i.e.{\ }within a
quasi-optimal realisation. However, the CRp-to-thermal energy density ratio
$X_\crp$ would have to compensate for this by increasing nearly two orders of
magnitude towards the outskirts of the halo.\footnote{Though, some part of this
  apparent increase is an artifact owing to azimuthally averaging the
  non-centrally symmetric synchrotron brightness distribution \citep[for a more
  detailed discussion on this topic, see][]{2004A&A...413...17P}.}  This choice
of the magnetic field morphology ($\alpha_B = 0.5$) has been adopted in
Fig.~\ref{fig:parameter} which represents a parameter study on the ability of
hadronically originating CRe to generate the radio halo of Coma.  Contour lines
of $\mbox{max}(X_\crp) = (1,0.3,0.1,0.03,0.01)$ for the range $r\le
1\,h_{70}^{-1}\mbox{ Mpc}$ are shown in parameter space spanned by $\alpha_\p$
and $B_0$. The gradient of the maximum of $X_\crp$ points downwards in
Fig.~\ref{fig:parameter} and thus leaves the upper region of parameter space
where the hadronic scenario is energetically able to account for the observed
radio halo. For the choice of $\alpha_\p=2.3$ and $X_B=0.01$ the maximum of the
CRp-to-thermal energy density $X_\crp$ is smaller than 3\% for the entire range
of the radio halo. Conservative choices for CRp spectral breaks have been
adopted by means of Eq.~(\ref{cutoffs}): We assume a high-momentum break of
$p_2 \sim 2\times 10^{7} \mbox{ GeV} c^{-1}$ being derived from CRp diffusion
\citep{1997ApJ...487..529B} while the lower momentum cutoff assumes the CRp to
be accelerated from a thermal Maxwellian distribution, $p_1 \sim 3 \sqrt{2 m_\p
  k T_\rmn{Coma}} = 0.01 \mbox{ GeV} c^{-1}$ \citep{2001CoPhC.141...17M}. This
choice of the lower cutoff represents the energetically tightest constraint
because taking into account Coulomb losses would only weaken the energetic
requirements.  Moving away from the minimum energy solution, especially in the
inner parts of the cluster, the presented energetic considerations show that
the hadronic scenario is a viable explanation of the Coma radio halo as long as
the spatially constant CRp spectral index is between $1.4 \lesssim \alpha_\p
\lesssim 2.8$.

\citet{1993ApJ...406..399G} found a strong radial spectral steepening from
$\alpha_\nu = 0.8 - 1.8$ which would translate within the hadronic scenario
into a CRp spectral index steepening of $\alpha_\p = 1.6 - 3.6$.  If the strong
steepening will be confirmed, the hadronic scenario will face serious
challenges even when including conservative CRp spectral breaks. However, an
absent or weaker steepening of the CRp spectral index e.g.{\ }from
$\alpha_\p=2.3$ in the cluster centre to $\alpha_\p=2.8$ at the outskirts of
the radio halo would only double the CRp energy density required to explain the
radio halo in the hadronic scenario. The studies of \citet{1993ApJ...406..399G}
are based on two synthesis aperture radio maps obtained with different radio
telescopes. The technique of interferometric radio observations generally
suffers from missing short-baseline information leading to an uncertainty of
emission from larger structures: the so-called ``missing zero
spacing''-problem.  This uncertainty of the surface brightness distribution at
a single frequency is even increased for spatial distributions of the spectral
index which represent a ratio of surface brightness distributions yielding to
possible observational artefacts at the outskirts of the radio halo.  Thus,
future observations are required to decide whether the strong spectral
steepening as a function of radius is an observational artifact or a real
characteristic of the radio halo.

\begin{figure}
\begin{tabular}{c}
\resizebox{\hsize}{!}{\includegraphics{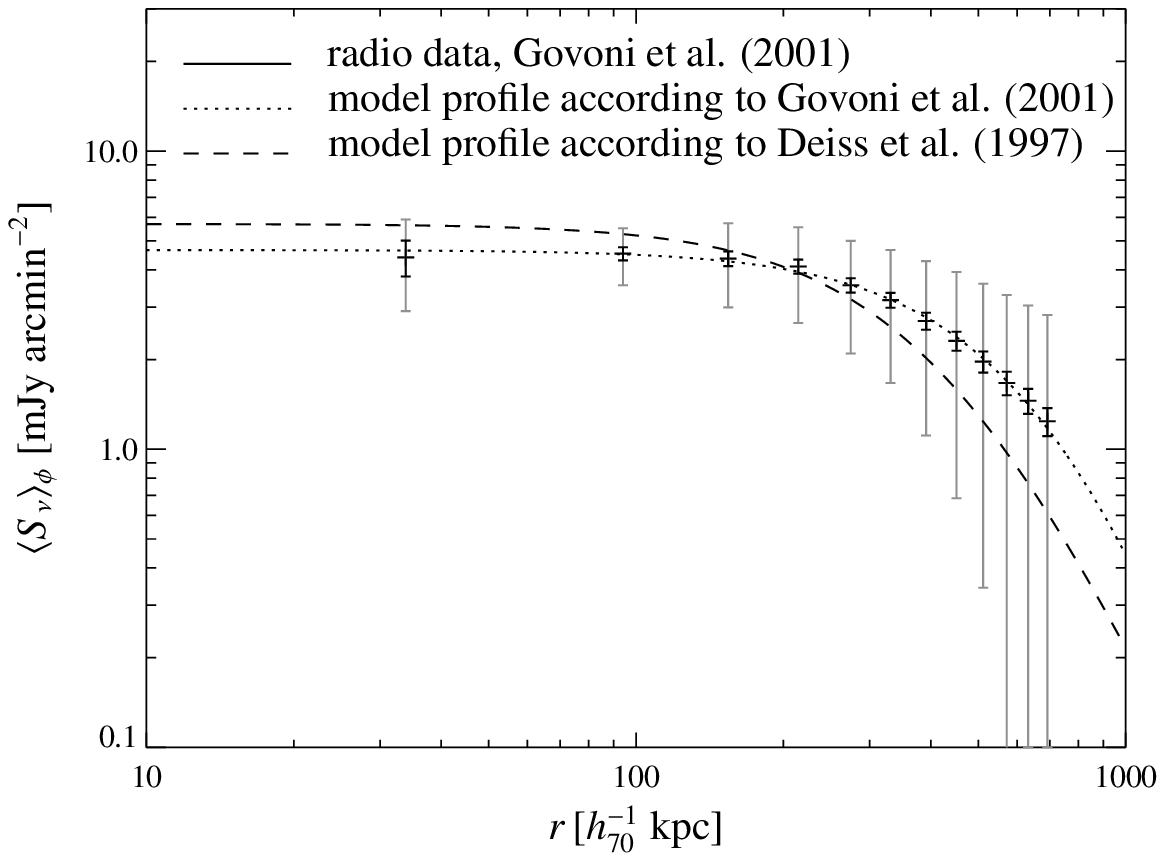}} \\
\\
\resizebox{\hsize}{!}{\includegraphics{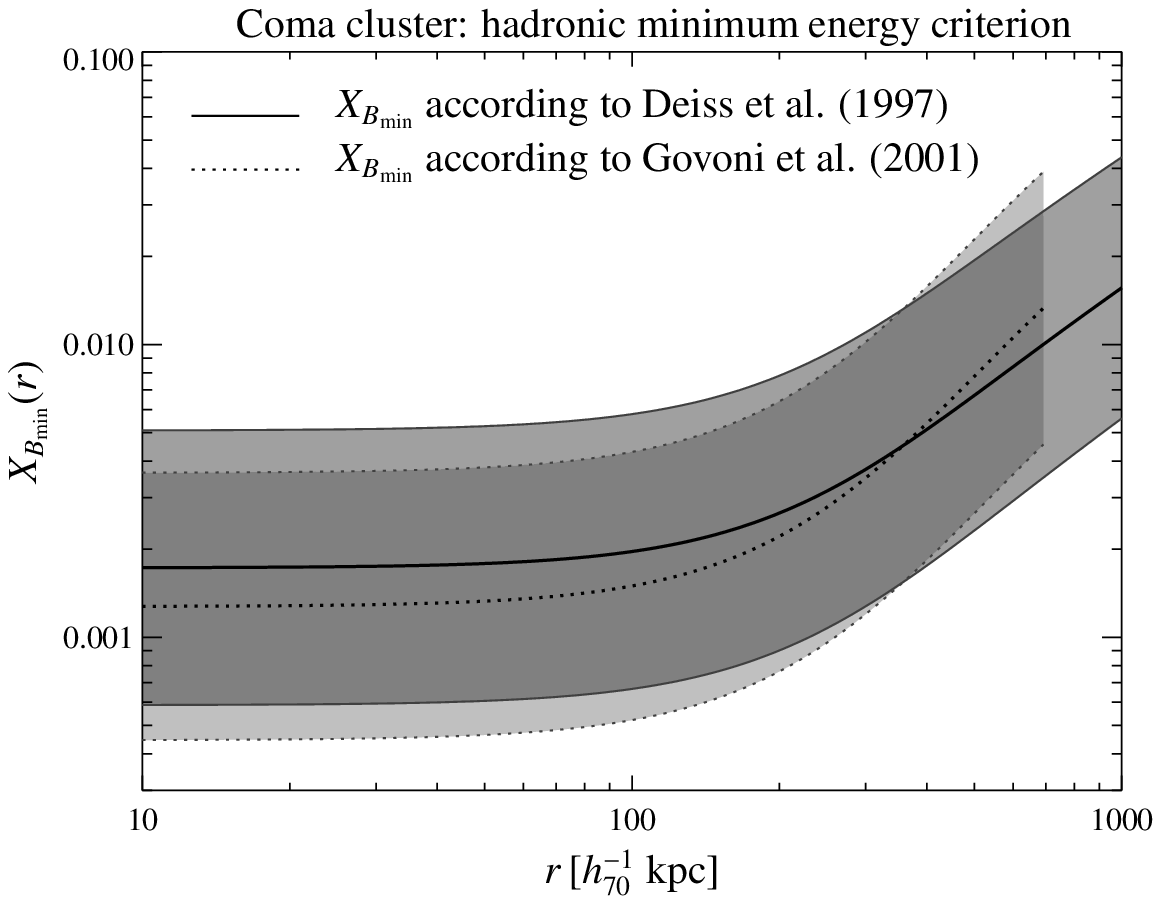}}
\end{tabular}
\caption{{\bf Top panel:} Azimuthally averaged radio brightness profile of
  the radio halo in the Coma cluster as a function of impact parameter
  $r_\bot$. Shown are the radio data at 326~MHz \citep{2001A&A...369..441G} in
  combination with the $1 \sigma$-error bars (black) and the surface brightness
  fluctuations within concentric annuli (grey) which are based on observations
  by \citet{1990AJ.....99.1381V}. Also presented is the model profile according
  to \citet{2001A&A...369..441G} (dotted) and the model profile according to
  \citet{1997A&A...321...55D} (dashed) which is rescaled to 326~MHz using the
  synchrotron spectral index $\alpha_\nu=1.15$ (c.f.{\ }Table
  \ref{tab:radioprofiles}). {\bf Lower panel:} Profiles of the
  magnetic-to-thermal energy density $X_{B_\rmn{min}}(r)$ in the Coma cluster
  as a function of deprojected radius are shown within the hadronic minimum
  energy criterion. A comparison of the used synchrotron profiles by
  \citet{1997A&A...321...55D} (solid, tolerance region medium grey shaded) and
  \citet{2001A&A...369..441G} (dotted, tolerance region light grey shaded)
  shows no significant difference within the allowed logarithmic tolerance
  regions (overlap is shown dark shaded).}
\label{fig:magComa}
\end{figure}

Figure \ref{fig:magComa} compares radio synchrotron profiles of the Coma radio
halo by \citet{2001A&A...369..441G} which is based on observations by
\citet{1990AJ.....99.1381V} using a synthesis aperture telescope with
observations by \citet{1997A&A...321...55D} using a single-dish telescope.  The
statistical variance given by \citet{2001A&A...369..441G} represents the rms
scatter within concentric annuli (shown in light grey) which is composed of
measurement uncertainties and non-sphericity of the underlying radio profile.
Rescaling with the square root of the number of independent beams within
concentric annuli yields statistical uncertainties (black) however without
taking into account systematics.\footnote{Due to the interferometric nature of
  the measurement and due to the non-completely synthesised aperture, we expect
  the true error bars to be larger than the estimates given here suggest.
  However, a detailed discussion of this topic is beyond the scope of this
  work.}  The top panel shows an comparison of the different profiles by
\citet{2001A&A...369..441G} and \citet{1997A&A...321...55D} (rescaled from its
original observational frequency $\nu = 1.4$~GHz to 326~MHz using the
synchrotron spectral index $\alpha_\nu=1.15$).  There seems to be an indication
of spectral steepening as reported by \citet{1993ApJ...406..399G}. However, for
simplicity we use a spatially constant CRp spectral index.  Because of the more
extended profile of the radio halo of the single-dish observation, we decided
to adopt the profile obtained by \citet{1997A&A...321...55D} in our analysis
shown in Fig.~\ref{fig:MEC}.  Nevertheless, the lower panel of
Fig.~\ref{fig:magComa} shows a comparison of the energetically favoured
magnetic-to-thermal energy density $X_{B_\rmn{min}}(r)$ as a function of
deprojected radius within the hadronic minimum energy criterion for both data
sets.  The tolerance regions of $X_{B_\rmn{min}}$ are drawn light shaded while
the overlap of $X_{B_\rmn{min}}$ using the different synchrotron profiles is
shown dark shaded. There is no significant difference within the allowed
tolerance regions.  Together with the previous considerations about spectral
steepening, this indicates that a moderate radially dependent spectral index
does not significantly modify the hadronic minimum energy condition while a
strong steepening would challenge the hadronic scenario.

% --- section: Minimum energy criterion in a nutshell --- %
\section{Minimum energy criteria in a nutshell}
\label{sec:nutshell}

This section provides self-consistent recipes for applying the classical and
hadronic minimum energy criterion in typical observational situations. We
present formulae for inferring magnetic field strengths solely as a function of
observed flux per frequency, $\mathcal{F}_\nu$, luminosity distance to the
galaxy cluster, $D$, extent of the cluster measured in core radius,
$r_\rmn{c}$, observed frequency, $\nu$, and spectral index of the diffuse
synchrotron emission, $\alpha_\nu$, where the emissivity scales as $j_\nu
\propto \nu^{-\alpha_\nu}$.

The omnidirectional (i.e.~integrated over $4\upi$ solid angle) luminosity per
frequency is given by the volume integral of the synchrotron emissivity,
$j_\nu$,
\begin{equation}
  \label{eq:luminosity}
  \mathcal{L}_\nu = 4\upi \int \dd V\, j_\nu.
\end{equation}
We choose a reference luminosity $\mathcal{L}_{\nu(1+z),0} =
4\upi\,D^2\,(1+z)^{-1}\,\mathcal{F}_{\nu,0}$ which corresponds to a
flux at $\nu = 1 ~\rmn{GHz}$ of $\mathcal{F}_{\nu,0} = 1~\rmn{Jy}$ for a source
at a luminosity distance of $D = 100~h_{70}^{-1}~\rmn{Mpc}$. This corresponds
to a cluster like Coma which is characterised by a core radius of $r_\rmn{c,0}
\sim 300~h_{70}^{-1}~ \rmn{kpc}$.

\subsection{Classical minimum energy criterion in a nutshell}

Applying the classical minimum energy criterion, we infer an optimal magnetic
field strength by rewriting Eq.~(\ref{eq:xB_min_class}),
\begin{eqnarray}
  \label{eq:Bclass}
  B_\rmn{min}^\rmn{class} &=& B_\rmn{min,0}^\rmn{class}(\alpha_\nu)~\umu\rmn{G}\,
  \left[\frac{1+k_\p}{2}\frac{\mathcal{L}_\nu}{\mathcal{L}_{\nu,0}}
    \left(\frac{r_\rmn{c}}{r_\rmn{c,0}}\right)^{-3}f_B^{-1}\right. 
\nonumber\\
  &&  \left.\times\,\left(\frac{\nu}{1~\rmn{GHz}}\right)^{\alpha_\nu}
    \left(\frac{E_1}{0.1~\rmn{GeV}}\right)^{1-2\alpha_\nu}
  \right]^{1/(\alpha_\nu + 3)}.
\end{eqnarray}
Here $B_\rmn{min,0}^\rmn{class}(\alpha_\nu)$ is given by Table~\ref{tab:nutshell}, $k_\p$
denotes the ratio between CRp and CRe energy densities, $E_1$ denotes the lower
cutoff of the CRe population, and $f_B$ denotes the filling factor of the
magnetic field in the volume occupied by CRe, which is thought to be of order
unity. While deriving Eq.~(\ref{eq:Bclass}) we implicitly assumed that $E_2 \gg
E_1$. We also applied a lower Coulomb cutoff to the CRe distribution of $E_1 =
0.1$~GeV \citep[as suggested by][]{1999ApJ...520..529S}, and adopt a
conservative choice for the cosmic ray energy scaling of $k_\p = 1$.

In the case of the classical minimum energy criterion the tolerance region of the
magnetic field strength is given by Eq.~(\ref{eq:sigma_B2}).  The following
substitutions might be useful when computing the tolerance levels of the
magnetic field, $\sigma_{\ln B} = \sigma_{\ln x_B}/2$, which are equally spaced
in logarithmic units of $B$. In linear representation of $B$, this definition
explicitly implies the tolerance levels given by $\exp(\ln B \pm
\sigma_{\ln B})$.  The scaled synchrotron index is given by $\delta =
(\alpha_\nu - 1)/2$, while the dimensionless magnetic energy density
$x_{B_\rmn{min}}^\rmn{class}$ and the constant $C_\rmn{class}(\alpha_\nu)$ are
denoted by
\begin{eqnarray}
  x_{B_\rmn{min}}^\rmn{class} &=& 
  \left(\frac{B_\rmn{min}^\rmn{class}}{B_\rmn{c}}\right)^2 = 
  \left(\frac{B_\rmn{min}^\rmn{class}}{31~\umu\rmn{G}}\right)^2
  \left(\frac{\nu}{1~\rmn{GHz}}\right)^{-2}, \\
  \label{eq:Cclassnutshell}
  C_\rmn{class}(\alpha_\nu) &=& f_\rmn{class}(\alpha_\nu)\,f_B^{-1}
  \frac{\mathcal{L}_\nu}{\mathcal{L}_{\nu,0}}
  \left(\frac{r_\rmn{c}}{r_\rmn{c,0}}\right)^{-3}
  \left(\frac{\nu}{1~\rmn{GHz}}\right)^{-3}, 
\end{eqnarray}
where $f_\rmn{class}(\alpha_\nu)$ is given by Table~\ref{tab:nutshell}.

\begin{table}
\caption[t]{Useful numerical values for particular choices of the synchrotron
  spectral index $\alpha_\nu$ in the framework of the minimum energy criterion
  in a nutshell are given for both scenarios. Note, that a priori, we assume no
  cutoff in the CRp distribution. For spectral indices $\alpha_\nu \le 1$
  within the hadronic scenario, an upper cutoff needs to be introduced for
  deducing an equipartition  magnetic field strength in order to meet the
  requirements of the regularity conditions.}   
\vspace{-0.1 cm}
\begin{center}
\begin{tabular}{cccccc}
\hline
$\alpha_\nu$ & $\alpha_\e$ & $B_\rmn{min,0}^\rmn{class}$ & $f_\rmn{class}$ 
                           & $B_\rmn{min,0}^\rmn{hadr}$  & $f_\rmn{hadr}$ \\
 & & $[\umu\rmn{G}]$ & & $[\umu\rmn{G}]$ & \\
\hline 
0.65 & 2.3 & 0.6 & $3.3 \times 10^{-7}$ & & \\
0.75 & 2.5 & 0.7 & $3.6 \times 10^{-7}$ & & \\
0.85 & 2.7 & 0.8 & $4.7 \times 10^{-7}$ & & \\
0.95 & 2.9 & 1.0 & $6.4 \times 10^{-7}$ & & \\
1.05 & 3.1 & 1.2 & $9.2 \times 10^{-6}$ & ~~4.0 & $2.1 \times 10^{-2}$ \\
1.15 & 3.3 & 1.4 & $1.3 \times 10^{-6}$ & ~~4.2 & $1.9 \times 10^{-2}$ \\
1.25 & 3.5 & 1.7 & $2.0 \times 10^{-6}$ & ~~5.3 & $3.4 \times 10^{-2}$ \\
1.35 & 3.7 & 2.0 & $3.0 \times 10^{-6}$ & ~~7.3 & $8.2 \times 10^{-2}$ \\
1.45 & 3.9 & 2.4 & $4.5 \times 10^{-6}$ &  13.0 & $3.9 \times 10^{-1}$ \\

\hline 
\end{tabular}
\end{center}
\label{tab:nutshell}
\end{table}

\subsection{Hadronic minimum energy criterion in a nutshell}

In the case of the hadronic scenario the energetically favoured magnetic field
strength $B_\rmn{min,0}^\rmn{hadr}$ is given by
\begin{equation}
  \label{eq:Bminhadr}
  B_\rmn{min,0}^\rmn{hadr} = \sqrt{x_{B_\rmn{min}}}\, B_\rmn{c} = 
  31~\umu\rmn{G}\, \sqrt{x_{B_\rmn{min}}}\, 
  \left(\frac{\nu}{1~\rmn{GHz}}\right),
\end{equation}
where $x_{B_\rmn{min}}$ is given by Eqs.~(\ref{eq:asymptotic}) through
(\ref{eq:asymptotic2}). $B_\rmn{min,0}^\rmn{hadr}$ is specified in
Table~\ref{tab:nutshell} for a few spectral indices $\alpha_\nu$ where we
assumed no cutoff of the CRp distribution. Provided the synchrotron index
$\alpha_\nu \le 1$, there must be an upper cutoff of the CRp distribution in
order to ensure a non-divergent CRp energy density. This might be obtained by
means of Eq.~(\ref{cutoffs}).  Owing to the non-analytic structure of the
hadronic minimum energy criterion (\ref{eq:Bfield}) in $x_{B_\rmn{min}}$, we
were forced to carry out an asymptotic expansion for $x_{B_\rmn{min}}$ which
does not admit a comparable simple scaling of the magnetic field as in the
classical case (\ref{eq:Bclass}).

The theoretically expected tolerance levels of the magnetic field,
$\sigma_{\ln B} = \sigma_{\ln x_B}/2$, are given by Eq.~(\ref{eq:sigma_B_hadr})
while neglecting the radial dependence in the nutshell approach.  The following
substitutions might be useful when computing the magnetic field strength
$B_\rmn{min,0}^\rmn{hadr}$ and the corresponding tolerance region. The scaled
synchrotron index is given by $\delta = (\alpha_\nu - 1)/2$, while the
dimensionless energy density of the CMB, $x_\rmn{CMB}$, and the constant
$C_\rmn{hadr}(\alpha_\nu)$ are denoted by
\begin{eqnarray}
  x_\rmn{CMB} &=& \frac{B_\rmn{CMB}^2}{B_\rmn{c}^2} = 
  1.08 \times 10^{-2}\,  \left(\frac{\nu}{1~\rmn{GHz}}\right)^{-2}(1+z)^4, \\
  \label{eq:Chadrnutshell}
  C_\rmn{hadr}(\alpha_\nu) &=& f_\rmn{hadr}(\alpha_\nu)\,f_B^{-1}
  \frac{\mathcal{L}_\nu}{\mathcal{L}_{\nu,0}}
  \left(\frac{r_\rmn{c}}{r_\rmn{c,0}}\right)^{-3}
  \left(\frac{n_\e}{n_{\e,0}}\right)^{-1}
  \left(\frac{\nu}{1~\rmn{GHz}}\right)^{-3}\!\!,
\end{eqnarray}
where $f_\rmn{hadr}(\alpha_\nu)$ is given by Table~\ref{tab:nutshell}, and
$n_{\e,0} = 10^{-3}~\rmn{cm}^{-3}$.

% --- section: Conclusions --- %
\section{Conclusions}

We investigated the minimum energy criterion of radio synchrotron emission in
order to estimate the energy density of magnetic fields with the main focus on
the underlying physical scenario.  The classical scenario might find
application for cosmic ray electrons (CRe) originating either from primary
shock acceleration or in-situ reacceleration processes while the hadronic
model assumes a scenario of inelastic cosmic ray proton (CRp) interactions
with the ambient gas of the intra-cluster medium (ICM) and thus leads to
extended diffuse synchrotron and $\gamma$-ray emission.

Generally, the hadronic minimum energy estimates allow testing the hadronic
model for extended radio synchrotron emission in clusters of galaxies.  If it
turns out that the required minimum non-thermal energy densities are too large
compared to the thermal energy density, the hadronic scenario has to face
serious challenges.  For the classical minimum energy estimate, such a
comparison can yield constraints on the accessible parameter space spanned by
the lower energy cutoff of the CRe population or the unknown contribution of
CRp to the non-thermal energy density.

For the first time we examine the localisation of the predicted minimum energy
densities and provide a measure of the theoretically expected tolerance regions
of these energetically favoured energy densities. The tolerance regions of the
particular energy densities inferred from the classical minimum energy
criterion are approximately constant for varying magnetic field strength
$\eps_B$. On the contrary, the hadronic minimum energy criterion predicts
constant energy densities for varying magnetic field strength in the case of
low $\eps_B$ compared to $\eps_\rmn{CMB}$, while the tolerance region of the
CRp energy density decreases at the same rate as the tolerance region of
$\eps_B$ increases for high $\eps_B$.

Future observations should shed light on the hypothetical realisation of such
an optimal distribution of energy densities in Nature: Combining upper limits
on the inverse Compton (IC) scattering of cosmic microwave background photons
off CRe within the ICM provides lower limits on the magnetic field
strength.  Unambiguous detection of the $\pi^0$-decay induced $\gamma$-ray
emission owing to hadronic CRp interactions in the ICM together with the
observed radio synchrotron emission yields strong upper limits on the magnetic
field strength. These are only upper limits because the inevitably accompanying
hadronically generated CRe could have a non-hadronic counterpart CRe population
which also contributes to the observed synchrotron emission. A combination of
IC detection in hard X-rays, radio synchrotron emission, and hadronically
induced $\gamma$-ray emission therefore simultaneously enables the
determination of the CRp population as well as a bracketing of the total
magnetic field strength and the CRe population. Applying the appropriate minium
energy arguments would yield information about both the dynamical state as well
as the fragmentation of the spatial distribution of the magnetic field.

Requiring the sum of cosmic ray and the magnetic field energy densities to be
minimal for the observed synchrotron emission of the radio halo of the Coma
cluster and the radio mini-halo of the Perseus cluster yields interesting
results: Within the theoretically expected tolerance regions, equipartition is
possible between the energy densities of CRp and magnetic fields, i.e.~the
minimum energy criterion always seems to choose equipartition to be a
quasi-optimal case.  Applying the hadronic minimum energy criterion to the
diffuse synchrotron emission of the Coma cluster yields a central magnetic
field strength of $B_\rmn{Coma} = 2.4^{+1.7}_{-1.0}~\umu\rmn{G}$ while in the
case of the cool core cluster Perseus we obtain $B_\rmn{Perseus} =
8.8^{+13.8}_{-5.4}~\umu\rmn{G}$.  These values agree with magnetic field
strengths inferred from Faraday rotation which range in the case of clusters
without cool cores within $[3~\umu\rmn{G}, 6~\umu\rmn{G}]$ while cool core
clusters yield values of $\sim 12~\umu\rmn{G}$ \citep{2003A&A...412..373V}.
Within the hadronic model for the radio mini-halo in the Perseus cluster, this
results in a confinement for the CRp energy density of $2\%\pm 1\%$ of the
thermal energy density while the magnetic energy density reaches only 0.4\% of
the thermal energy density within large uncertainties.  These energetic
considerations show that the hadronic scenario is a very attractive explanation
of cluster radio mini-halos.

In order to account for the radio halo of Coma in the hadronic scenario, the
product of $\eps_\crp$ and $\eps_B$ needs to increase by nearly two orders of
magnitude relative to the square of the thermal energy density $\eps_\rmn{th}$
towards the outskirts of the halo.  Moving away from the minimum energy
solution and adopting for instance a constant magnetic-to-thermal energy
density, it is energetically possible to explain the observed synchrotron
emission hadronically by only requiring the magnetic and CRp energy density to
be a few per cent relative to the thermal energy density (and even less for the
CRp in the cluster centre, provided $\alpha_\p \sim 2.3$ and the cluster is
isothermal). Such a magnetic energy density corresponds to a central magnetic
field strength of $6~\umu$G. Assuming a lower magnetic field strength of
$3~\umu$G corresponding to a magnetic-to-thermal energy density of
approximately 0.5\% requires the CRp energy density to be lower than 10\% for the
entire range of the radio halo.
  
The considered hadronic scenario assumes a CRp spectral index which is
independent of position and thus the radio emission does not show any spatial
variations over the clusters volume. In principle, one could allow for radial
spectral variations of the CRp and thereby for the radio emission by adopting a
particular history of this population.  For instance, one possible scenario
would be given by continuous in-situ acceleration of CRp via resonant pitch
angle scattering by turbulent Alfv\'en waves. We discuss that a moderate radial
steepening would not significantly modify the hadronic minimum energy condition
while a confirmation of the strong steepening reported by
\citet{1993ApJ...406..399G} would seriously challenge the hadronic scenario.

As a caveat, it should be stressed that the inferred values for the particular
energy densities only represent the energetically least expensive radio
synchrotron emission model possible for a given physically motivated scenario.
This minimum is not necessarily realised in Nature. Nevertheless, our
minimum energy estimates are also interesting in a dynamical respect: Should
the hadronic scenario of extended radio synchrotron emission be confirmed, the
minimum energy estimates allow testing for the realisation of the minimum
energy state for a given independent measurement of the magnetic field
strength.  Within the tolerance regions, our minimum energy estimates in
Perseus and Coma agree well with magnetic field strengths inferred from Faraday
rotation.  Under the hypotheses of correctness of the hadronic scenario, such a
possible realisation of the minimum energy state would seek an explanation of a
first principle enforcing this extremal value to be realised in Nature.

% --- section: --- %
\section*{Acknowledgements}
The authors would like to thank Simon D.~M.~White, Matthias Bartelmann, Philip
P.~Kronberg, Bj\"orn Malte Sch\"afer, Corina Vogt and an anonymous referee for
carefully reading the manuscript and their constructive remarks.  This work was
performed within the framework of the European Community Research and Training
Network {\it The Physics of the Intergalactic Medium}.

% --- section: bibliography --- %
\bibliography{bibtex/chp} 
\bibliographystyle{mn2e}

\appendix

\bsp

\label{lastpage}

\end{document}